\documentclass[a4paper,11pt]{article}
\usepackage{physics}
\usepackage{jcappub}
\usepackage{pgfplots}

\usepackage{amsmath}
\usepackage{bm}
\usepackage{xcolor}
\usepackage{float}
\usepackage{multirow}
\usepackage{makecell}

\newcommand{\aap}{{Astron.~Astrophys.}}
\newcommand{\apjl}{{Astrophys.~J.~Lett.}}
\newcommand{\apjs}{{Astrophys.~J.~Supp.}}

\usepackage{graphicx}
\usepackage{subcaption}
\usepackage{amssymb, amsmath}
\usepackage[amssymb]{SIunits}
\usepackage{epstopdf}
\usepackage{hyperref}
\usepackage{aas_macros}

\usepackage[mathscr]{eucal}	
\DeclareTextSymbol{\degre}{T1}{23}

\newcommand{\rmd}{\mathrm{d}}

\newcommand{\beq} {\begin{equation}}
\newcommand{\eeq} {\end{equation}}
\newcommand{\bal} {\begin{aligned}}
\newcommand{\eal} {\end{aligned}}

\newcommand{\refeq}[1]{Eq.~(\ref{eq:#1})}          
          
\newcommand{\reffig}[1]{Figure~\ref{fig:#1}}          
\newcommand{\refsec}[1]{Section~\ref{sec:#1}}

\newcommand{\reftab}[1]{Table~\ref{tab:#1}}

\newcommand{\ie}{\emph{i.e.}~}

\newcommand{\lf}{\left}
\newcommand{\ri}{\right}
\newcommand{\vect}[1]{\boldsymbol{\mathbf{#1}}} 
\newcommand{\hatv}[1]{\boldsymbol{\mathbf{\hat{#1}}}} 

\title{Probing cosmic strings by reconstructing polarization rotation of the cosmic microwave background}

\author[a]{Weichen Winston Yin,}
\author[a]{Liang Dai,}
\author[b, a]{and Simone Ferraro}
\affiliation[a]{Department of Physics, 366 Physics North MC 7300, University of California, Berkeley, CA 94720, USA}
\affiliation[b]{Lawrence Berkeley National Laboratory, One Cyclotron Road, Berkeley, CA 94720, USA}
\emailAdd{winstonyin@berkeley.edu}
\emailAdd{liangdai@berkeley.edu}
\emailAdd{sferraro@lbl.gov}

\abstract{
Cosmic birefringence---the rotation of the polarization of the cosmic microwave background (CMB) photons as they travel to the Earth---is a smoking gun for axion-like particles (ALPs) that interact with the photon. It has recently been suggested that topological defects in the ALP field called cosmic strings can cause polarization rotation in quantized amounts that are proportional to the electromagnetic chiral anomaly coefficient $\mathcal A$, which holds direct information about physics at very high energies. In this work, we study the detectability of a random network of cosmic strings through estimating rotation using quadratic estimators (QEs). We show that the QE derived from the maximum likelihood estimator is equivalent to the recently proposed global-minimum-variance QE; the classic Hu-Okamoto QE equals the global-minimum-variance QE under special conditions, but is otherwise still nearly globally optimal. We calculate the sensitivity of QEs to cosmic birefringence from string networks, for the Planck satellite mission, as well as for third- and fourth-generation ground-based CMB experiments. Using published binned rotation power spectrum derived from the Planck 2015 polarization data, we obtain a constraint $\mathcal A^2\,\xi_0 < 0.93$ at the 95\% confidence level, where $\xi_0$ is the total length of strings in units of the Hubble scale per Hubble volume, for a phenomenological but reasonable string network model describing a continuous distribution of string sizes. Looking forward, experiments such as the Simons Observatory and CMB-S4 will either discover or falsify the existence of an ALP string network for the theoretically plausible parameter space $\mathcal A^2\,\xi_0 \gtrsim 0.01$.
}

\begin{document}

\maketitle


\section{Introduction}\label{sec:introduction}

The intensity and polarization anistropies of the cosmic microwave background (CMB) are the most stringently tested physical observables in cosmology. At the leading order, the anisotropies are Gaussian random quantities and follow isotropic power spectra that are precisely predictable from physical models of the primordial Universe. Nearly originated from the cosmic horizon, the CMB photons traverse the longest distances in the Universe without major absorption or scattering.

The above two reasons render the CMB a sensitive probe of cosmic birefringence---the coherent rotation of the polarization planes as CMB photons travel along the line of sight. This effect can be mediated by an ultralight axion-like particles (ALPs, or simply ``axions'' in this work) that couples to the photon via a Chern-Simons coupling \cite{Carroll_1998, CarrollFieldJackiw_1990, HarariSikivie_1992}. Such ALPs are a ubiquitous outcome of string theory compactifications \cite{SvrcekWitten_2006_StringTheory}; they may account for the dark matter \cite{Hui_2017}, or hold clues to the mystery of the Dark Energy \cite{Kamionkowski2014axiverse, Razieh2016axiverse}. Recently, Ref.~\cite{Agrawal_2019CMBmillikan} pointed out that topological defects in the ALP field, termed cosmic strings, can imprint birefringence patterns in the CMB in a quantized manner. Remarkably, the observable rotation angle is insensitive to the ALP-photon coupling constant, which is unconstrained and subject to renormalization. In fact, the rotation angle varies in quantized amounts as the line of sight crosses individual strings, which is determined by the electromagnetic chiral anomaly coefficient $\mathcal{A}$, free from renormalization \cite{Hooft1980}. Not only does $\mathcal A$ offer insight into the smallest unit of electric charge \cite{Witten:1979ey}, it is also useful for distinguishing between beyond-the-Standard-Model theories, whose matter content and charge assignment at very high energies lead to predictions for the value of $\mathcal A$ \cite{Agrawal_2019CMBmillikan}.

Recently, the study of CMB birefringence has heated up after tentative evidence for birefringence that is highly coherent across the sky was reported in a refined analysis of the Planck 2018 polarization data \cite{Minami_2020}, with implications for ALP searches \cite{Fujita_2021}. While an isotropic polarization rotation would be evidence for symmetry-breaking phenomena beyond the Standard Model, it is the power spectrum of direction-dependent rotations that will inform us about the spatial structure of birefringence sources.

While the anisotropic rotation induced by the strings is non-Gaussian by nature, in this work we study detectability using quadratic estimators which are optimal for Gaussian random fields. In this method, one seeks statistical correlation between Fourier modes of the anisotropies with different Fourier wave vectors. This method is appropriate for the scenario of a large number of strings within the observable Universe, but should also offer decent detection statistics in the case of relatively few strings. This simplification may be adequate if one aims for a first statistical detection of rotation, before characterization of the non-Gaussian behaviors of the rotation pattern is carried out.

Quadratic estimators for the purpose of estimating direction-dependent birefringence were first investigated in Refs.~\cite{Kamionkowski_2009, Gluscevic2009Derotate} in the full-sky formalism and in Ref.~\cite{Yadav_2009} in the flat-sky approximation. This method was used to derive bounds on the rotation power spectrum in the WMAP data~\cite{Gluscevic2012WMAP7}. In this work, we compare a few differently constructed quadratic estimators in general as well as in the context of cosmic birefringence. We study the detectability of direction-dependent birefringence as a result of a network of ALP cosmic strings \cite{Agrawal_2019CMBmillikan, Jain_2021}, for CMB anisotropy data collected by the Planck mission as well as forthcoming data to be collected by third and fourth generation ground-based CMB experiments. Instead of using the $EB$ and/or $TB$ estimators alone (as in Refs.~\cite{Yadav_2009} and \cite{Agrawal_2019CMBmillikan}), we maximize information gain by including all quadratic combinations of CMB anisotropies. We give signal-to-noise ratio (SNR) forecasts for measuring the parameter combination $\mathcal A^2\,\xi_0$ through these CMB experiments, where $\xi_0$ is the total length of strings per Hubble volume in units of the Hubble length, as described in \refsec{models}.

In the absence of cosmic birefringence, the intervening large-scale structure of the Universe induces cross-correlation between Fourier modes of the CMB anisotropies via gravitational lensing. Nevertheless, we will show that estimation of polarization rotation is nearly orthogonal to the estimation of the lensing potential, as was also discussed in Ref.~\cite{Gluscevic2009Derotate}. Therefore, for the purpose of detecting polarization rotation, it is a good approximation to treat the lensed anistropies as Gaussian random fields with statistical isotropy and to completely neglect mode correlations induced by lensing.

We study three ways to construct quadratic estimators for the rotation field using primary anisotropies. The first one is the famous Hu-Okamoto (HO) estimator \cite{Hu_2002}, which has been widely employed in weak lensing reconstruction of the CMB. In the second way, the so-called global-minimum-variance (GMV) estimator is constructed following the recent work of Ref.~\cite{Maniyar_2021}, which improves upon the HO estimator. For the third one, we derive an optimal quadratic estimator following the maximization of the full likelihood function for the anisotropy observables.

We explicitly show that the third quadratic estimator, obtained as an approximation to the maximal likelihood estimator, is equivalent to the GMV estimator, as one would anticipate based on the definition of the GMV estimator. Although the GMV estimator is in general distinct from the HO estimator and is mathematically guaranteed to be at least as sensitive as the latter \cite{Maniyar_2021}, we show that the two are equivalent when rotation estimators are constructed only from the $E$- and $B$-mode polarization anisotropies, which includes the dominant $EB$ estimators. We phrase a general sufficient condition for such equivalence. Furthermore, we numerically find that if the temperature anisotropies are also included, then the difference in the 
reconstruction noise between the GMV and HO estimators for the rotation field is insignificant at both current and forthcoming CMB experiments, with less than 0.5\% improvement from GMV across the range of angular scales.

We find that information about the plausible cosmic string network is dominated by rotation reconstruction at large scales $L \lesssim 100$, and that data from the Planck mission should already have constraining power for a significant part of the theoretically favored parameter space, $\mathcal A^2\,\xi_0 \gtrsim \mathcal{O}(1)$. For the entire theoretically plausible parameter range $\mathcal A^2\,\xi_0 \gtrsim 0.01$, we find that future CMB experiments will have the definitive statistical power to either substantiate or falsify the existence of cosmic strings from ALPs that couple to electromagnetism.

Using published birefringence power spectrum extracted from Planck 2015 data \cite{planck_constraint_2017}, we place upper bounds on the parameter $\mathcal A^2\,\xi_0$. For two phenomenological string network models developed in \cite{Jain_2021}, we find that $\mathcal A^2\,\xi_0 < 8.0$ if the Universe is populated by string loops of the same but arbitrary size, and $\mathcal A^2\,\xi_0 < 0.93$ if, more realistically, the string loops have a continuous distribution of sizes, both at the 95\% confidence level. These results are consistent with the absence of a cosmic string network.

This paper is organized as follows. \refsec{formalism} reviews the construction of the HO and GMV estimators. In \refsec{max_likelihood}, we derive a quadratic estimator from the maximum likelihood method. In \refsec{relation}, we show that the maximum likelihood quadratic estimator is equivalent to the GMV estimator; in addition, HO and GMV estimators constructed from only the polarization anisotropies are equivalent, and a general sufficient condition for such equivalence is stated. In \refsec{computation}, we numerically compute the reconstruction noise spectra for the HO and GMV estimators, respectively, showing that the difference is insignificant in practice. In \refsec{models}, we compute the overall signal-to-noise ratio for inferring the existence of a cosmic string network. We show in \refsec{simultaneous} that the bias induced by lensing on estimating rotation is at most second order in the lensing potential, justifying the approximation of lensed CMB anisotropies as Gaussian random fields. Finally in \refsec{bounds}, birefringence power spectra extracted from Planck 2015 is used to constrain parameters in string network models. Concluding remarks are given in \refsec{conclusion}.

\section{Quadratic estimators}
\label{sec:formalism}

As the CMB photons travel from the last scattering surface, they follow deflected trajectories due to gravitational deflection by matter overdensity and underdensity. Along the perturbed trajectories, the axion-photon coupling causes a rotation of the plane of polarization. At the leading order, it suffices to compute the rotation effect along the unperturbed line of sight. The absence of interference between these two effects at the leading order is discussed and justified in \refsec{simultaneous}. As we focus on the rotation effect, we start with the {\it lensed but unrotated} CMB anisotropies, and consider the effect of rotation on them.

\subsection{Notation}

Throughout this paper, we assume a flat-sky approximation, in which the observables are functions of $\hatv n = n_x\,\hatv x + n_y\,\hatv y$. We use the tilded notation to indicate lensed but unrotated quantities: the temperature anisotropies $\tilde T(\hatv n)$, and the polarization anisotropies $\tilde E(\hatv n)$ and $\tilde B(\hatv n)$. Their power spectra in Fourier space $\tilde C^{XY}_{l}$ are defined through:
\begin{equation}
    \ev{\tilde X(\vect l_1)\tilde Y^*(\vect l_2)} = (2\pi)^2\,\delta_D(\vect l_1-\vect l_2)\,\tilde C^{XY}_{l_1}.
\label{eq:unrotated_power_spectra}
\end{equation}
We parameterize cosmic birefringence using an anisotropic rotation angle for the polarization $\alpha(\hatv n)$. The lensed and rotated quantities are denoted by $T(\hatv n)$, $E(\hatv n)$, and $B(\hatv n)$ (birefringence leaves the temperature anisotropies unchanged). The cross correlations are consequently modified to
\begin{equation}
    \ev{X(\vect l_1)Y^*(\vect l_2)} = \ev{\tilde X(\vect l_1)\tilde Y^*(\vect l_2)} + f_{XY}(\vect l_1,\vect l_2)\,\alpha(\vect l_1-\vect l_2) + O(\alpha^2).
\label{eq:linear_response}
\end{equation}
Based on the latest constraints from the CMB data \cite{Contreras_2017} and predictions of theoretical models \cite{Jain_2021}, we expect $\alpha(\hatv n)$ to be small angles (in radians). Therefore, it suffices to expand the expression up to the linear order.

The observed anisotropies (denoted with a subscript $o$) include both the cosmological signals and noise (denoted with a subscript $n$):
\begin{equation}
    X_o(\vect l) = X(\vect l) + X_n(\vect l).
\end{equation}
The Fourier space power spectra can be separately defined for the observed anisotropies and for the noise, respectively:
\begin{align}
    \ev{X_o(\vect l_1)Y_o^*(\vect l_2)} &= (2\pi)^2\,\delta_D(\vect l_1-\vect l_2)\,C^{XY}_{l_1}\\
    \ev{X_n(\vect l_1)Y_n^*(\vect l_2)} &= (2\pi)^2\,\delta_D(\vect l_1-\vect l_2)\,C^{XY,n}_{l_1}.
\end{align}
In this paper, we concern the variance of rotation estimators under the null hypothesis, \ie $\alpha(\hatv n) = 0$. Since cosmic signals and noise are uncorrelated, we have
\begin{equation}
    C^{XY}_l = \tilde C^{XY}_l + C^{XY,n}_l.
\end{equation}

\subsection{Rotation of polarization}

Let $\tilde Q(\hatv n)\pm i\tilde U(\hatv n)$ be the Stokes' parameters for the lensed but unrotated polarizations. For a line-of-sight specific polarization rotation $\alpha(\vect n)$, the perturbed Stokes fields are given by:
\begin{equation}
    Q(\hatv n)\pm iU(\hatv n) = \left[ \tilde Q(\hatv n)\pm i\tilde U(\hatv n) \right]\,e^{\pm 2\,i\,\alpha(\hatv n)},
\label{eq:rotation_action}
\end{equation}
The $E$- and $B$-modes can be derived from the Stokes' parameters via \cite{Seljak:1998md}
\begin{equation}
    E(\vect l) \pm iB(\vect l) = \int \rmd^2\hatv n\,e^{-i\vect l\cdot\hatv n}\, e^{\mp 2\,i\,\varphi_{\vect l}}\,[Q(\hatv n) \pm iU(\hatv n)],\label{EB_from_QU}
\end{equation}
where $\varphi_{\vect l}$ is the angle between the vector ${\vect l}$ and the first coordinate axis on the sky. A similar expression can be written for the lensed but unrotated anisotropies.

We have assumed that all primary CMB anisotropies originate from the last scattering surface, neglecting the possibility that the effects of the ALP field may be different for polarizations produced at recombination and at reionization. Future CMB satellites such as LiteBIRD will be able to detect any difference in the birefringence angle between the two epochs to within $0.05^\circ$ \cite{sherwin2021cosmic}. For the experiments we consider, this is much smaller than the reconstruction noise at all but the lowest few $L$ modes (see \reffig{noise_curves}). Due to the limited number of modes, reionization polarization signals are not expected to affect the overall detectability for even the lowest $L$ modes, regardless of the correct treatment of the rotation effect up to the epoch of reionization. Having the small-scale primary anisotropies available is sufficient for reconstructing the rotation on large angular scales.

In \reftab{response_funcs}, we summarize the response functions (introduced in \refeq{linear_response}) due to the action of rotation.

\begin{table}[h!]
    \centering
    \begin{tabular}{|c|c|}
        \hline
        $XY$ & $f_{XY}(\vect l_1,\vect l_2)$\\
        \hline
        $TT$ & $0$\\
        $TE$ & $-2\tilde C_{l_1}^{TE}\sin 2\varphi_{12}$\\
        $TB$ & $2\tilde C_{l_1}^{TE}\cos 2\varphi_{12}$\\
        $EE$ & $-2\lf(\tilde C_{l_1}^{EE} - \tilde C_{l_2}^{EE}\ri)\sin 2\varphi_{12}$\\
        $EB$ & $2\lf(\tilde C_{l_1}^{EE} - \tilde C_{l_2}^{BB}\ri)\cos 2\varphi_{12}$\\
        $BB$ & $-2\lf(\tilde C_{l_1}^{BB} - \tilde C_{l_2}^{BB}\ri)\sin 2\varphi_{12}$\\
        \hline
    \end{tabular}
    \caption{Response functions under the action of line-of-sight dependent rotation of polarization. Here we have defined the angle $\varphi_{12} = \varphi_{\mathbf l_1} - \varphi_{\mathbf l_2}$.}
    \label{tab:response_funcs}
\end{table}

\subsection{Hu-Okamoto estimator}

Following the treatment of Hu \& Okamoto for estimating weak lensing from the CMB anisotropies~\cite{Hu_2002}, quadratic estimators can be constructed from the observed quantities~\cite{Yadav_2009}:
\begin{equation}
    \hat\alpha_{XY}(\vect L) = \int\frac{\rmd^2\vect l_1\,\rmd^2\vect l_2}{(2\pi)^2}\,\delta_D(\vect l_1-\vect l_2-\vect L)\,X_o(\vect l_1)\,Y_o^*(\vect l_2)\,F_{XY}(\vect l_1,\vect l_2),\label{eq:alpha_XY}
\end{equation}
where $F_{XY}$ is a yet undetermined real-valued function,

and we have assumed $\vect L\neq 0$. The $\vect L=0$ case corresponds to a constant rotation of the polarization on the whole sky. See Refs.~\cite{Hinshaw_2013,Planck_2016,Polarbear_2020,Bianchini_2020,Namikawa_2020,Choi_2020,Minami_2020} for recent estimates. Henceforth, we will write $\vect L = \vect l_1-\vect l_2$. Following Ref.~\cite{Maniyar_2021}, we also introduce the shorthand notation:
\begin{equation}
     \int\frac{\rmd^2\vect l_1\,\rmd^2\vect l_2}{(2\pi)^2}\,\delta_D(\vect l_1-\vect l_2-\vect L) \to \int_{\vect L=\mathbf l_1-\mathbf l_2}.
\end{equation}

To find the coefficients $F_{XY}(\vect l_1,\vect l_2)$, three conditions need to be imposed on $\hat\alpha_{XY}(\vect L)$. First, the reality condition $\hat\alpha_{XY}^*(\vect L) = \hat\alpha_{XY}(-\vect L)$ is needed because the rotation field to be estimated is real-valued. This requires $F_{XY}(-\vect l_1,-\vect l_2) = F_{XY}(\vect l_1,\vect l_2)$. Second, the estimator is required to be unbiased, $\ev{\hat\alpha_{XY}(\vect L)} = \alpha(\vect L)$, which leads to equation
\begin{equation}
    1 = \int_{\vect L=\mathbf l_1-\mathbf l_2} f_{XY}(\vect l_1,\vect l_2)\,F_{XY}(\vect l_1,\vect l_2).
\label{eq:unbiased}
\end{equation}
Finally, subject to the above two conditions, we seek to minimize the variance of the estimator via the following condition:
\begin{equation}
    \frac{\delta}{\delta F_{XY}}\lf[\ev{\hat\alpha_{XY}(\vect L)\,\hat\alpha_{XY}^*(\vect L')} - \ev{\hat\alpha_{XY}(\vect L)}\ev{\hat\alpha_{XY}^*(\vect L')}\ri] = 0.
\end{equation}
The solution to this minimization problem reads:
\begin{align}
    F_{XY}(\vect l_1,\vect l_2) &= \lambda_{XY}(\vect L)\,\frac{f_{XY}(\vect l_1,\vect l_2)}{(1+\delta_{XY})\,C_{l_1}^{XX}\,C_{l_2}^{YY}}\quad\text{for $XY\neq TE$},\label{eq:F_XY}\\
    F_{TE}(\vect l_1,\vect l_2) &= \lambda_{TE}(\vect L)\,\frac{C^{EE}_{l_1}\,C^{TT}_{l_2}\,f_{TE}(\vect l_1,\vect l_2)-C^{TE}_{l_1}\,C^{TE}_{l_2}\,f_{TE}(\vect l_2,\vect l_1)}{C^{TT}_{l_1}\,C^{EE}_{l_2}\,C^{EE}_{l_1}\,C^{TT}_{l_2}-(C^{TE}_{l_1}\,C^{TE}_{l_2})^2},
\label{eq:F_TE}
\end{align}
where we define the normalization constants
\begin{align}
    [\lambda_{XY}(\vect L)]^{-1} &= \int_{\vect L=\mathbf l_1-\mathbf l_2} \frac{[f_{XY}(\vect l_1,\vect l_2)]^2}{(1+\delta_{XY})\,C^{XX}_{l_1}\,C^{YY}_{l_2}}\quad\text{for $XY\neq TE$}, \label{eq:lambda_XY}\\
    [\lambda_{TE}(\vect L)]^{-1} &= \int_{\vect L=\mathbf l_1-\mathbf l_2} f_{TE}(\vect l_1,\vect l_2)\, \frac{C^{EE}_{l_1}\,C^{TT}_{l_2}\,f_{TE}(\vect l_1,\vect l_2)-C^{TE}_{l_1}\,C^{TE}_{l_2}\,f_{TE}(\vect l_2,\vect l_1)}{C^{TT}_{l_1}\,C^{EE}_{l_2}\,C^{EE}_{l_1}\,C^{TT}_{l_2}-(C^{TE}_{l_1}\,C^{TE}_{l_2})^2}. \label{eq:lambda_TE}
\end{align}
The covariance between two estimators is
\begin{equation}
    \ev{\hat\alpha_{XY}(\vect L)\hat\alpha_{X'Y'}^*(\vect L')} - \ev{\hat\alpha_{XY}(\vect L)}\ev{\hat\alpha_{X'Y'}^*(\vect L')} = (2\pi)^2\delta(\vect L-\vect L')N_{XYX'Y'}(\vect L),
\end{equation}
where
\begin{equation}
    N_{XYX'Y'}(\vect L) = \int_{\vect L=\vect l_1-\vect l_2} F_{XY}(\vect l_1,\vect l_2)\left[F_{X'Y'}(\vect l_1,\vect l_2)C^{XX'}_{l_1}C^{YY'}_{l_2} + F_{X'Y'}(\vect l_2,\vect l_1)C^{XY'}_{l_1}C^{YX'}_{l_2}\right]. \label{eq:covariance}
\end{equation}
The estimator constructed for any single pair $XY$ has a variance
\begin{equation}
    \ev{\hat\alpha_{XY}(\vect L)\,\hat\alpha_{XY}^*(\vect L')} - \ev{\hat\alpha_{XY}(\vect L)}\ev{\hat\alpha_{XY}^*(\vect L')} = (2\pi)^2\,\delta_D(\vect L-\vect L')\,N_{XY}(\vect L),
\label{eq:variance}
\end{equation}
where it can be shown that $N_{XY}(\vect L) = N_{XYXY} = \lambda_{XY}(\vect L)$. We note that $N_{XY}(\vect L)$ is only dependent on $L=|{\vect L}|$ due to statistical invariance under rotation of the sky. From here on, we will write $N_{XY}(L)$ to reflect this fact.

The Hu-Okamoto (HO) estimator is constructed as the unique linear combination of the estimators $\hat\alpha_{XY}(\vect L)$ corresponding to five pairs, $TE$, $TB$, $EE$, $EB$, and $BB$ (the $TT$ estimator does not exist, as pointwise polarization rotations have no effect on the temperature anisotropies), such that the linear combination remains unbiased and has a minimized variance:
\begin{equation}
    \hat\alpha_{HO}(\vect L) = \sum_I\,w_I(L)\,\hat\alpha_I(\vect L),
\label{eq:HO}
\end{equation}
where the index $I$ ranges over unique pairs of observables $XY$. Defining the symmetric matrix $\vect N(L)$ with elements $[\vect N]_{IJ}(L) = N_{IJ}(L)$ as defined in \refeq{covariance}, the weights can be computed from
\begin{equation}
    w_I(L) = \frac{\sum_J [\vect N^{-1}(L)]_{IJ}}{\sum_{JK}[\vect N^{-1}(L)]_{JK}}.\label{eq:weights}
\end{equation}
Under the null hypothesis that there is no polarization rotation, the HO estimator has a variance
\begin{equation}
    \ev{\hat\alpha_{HO}(\vect L)\,\hat\alpha_{HO}^*(\vect L')} - \ev{\hat\alpha_{HO}(\vect L)}\ev{\hat\alpha_{HO}^*(\vect L')} = (2\pi)^2\,\delta_D(\vect L-\vect L')\,N_{HO}(\vect L),
\label{eq:HO_variance}
\end{equation}
where
\begin{equation}
    N_{HO}(\vect L) = N_{HO}(L) = \left( \sum_{IJ} [\vect N^{-1}(L)]_{IJ} \right)^{-1}.
\label{eq:noise_HO}
\end{equation}
We note that the $5\times 5$ symmetric matrix $\vect N(L)$ is block diagonal; it is comprised of the $1\times 1$ block for $BB$, the $2\times 2$ block for $TE$ and $EE$, and the $2\times 2$ block for $TB$ and $BB$. This greatly simplifies the inversion of the matrix.

\subsection{Global-minimum-variance estimator}\label{sec:GMV}

Recently in the context of weak lensing reconstruction, Ref.~\cite{Maniyar_2021} proposes another quadratic estimator called the global-minimum-variance (GMV) estimator, which is directly constructed from a minimum-variance linear combination of pairs of {\it all} observables. This idea can be straightforwardly applied to the case of cosmic birefringence, leading to the following quadratic estimator
\begin{equation}
    \hat\alpha_{GMV}(\vect L) = \int_{\vect L=\mathbf l_1-\mathbf l_2} X^i_o(\vect l_1)\,X^{j*}_o(\vect l_2)\,\Xi_{ij}(\vect l_1,\vect l_2), 
\label{eq:GMV}
\end{equation}
where Einstein summation is assumed over $i$ and $j$, and $X^i = T$, $E$, or $B$. In contrast, the HO estimator, discussed in the previous subsection and historically widely used, is constructed as the minimum-variance linear combination of estimators which are in turn constructed from pairs of observables that correspond to different Fourier vectors. The work of Ref.~\cite{Maniyar_2021} shows that the reconstruction noise (variance) of the HO estimator is strictly no less than that of the GMV estimator. Hence, the HO estimator is a sub-optimal quadratic estimator. The two-step process to construct the HO estimator compromises the optimality. In fact, the GMV estimator achieves the least variance among all unbiased quadratic estimators.

The derivation needed to specify the GMV estimator closely parallels that in the previous subsection. Here, we summarize the relevant results following the notation of Ref.~\cite{Maniyar_2021}. To this end, we introduce bold face symbols, $\vect f$, $\vect C_l$, and $\vect\Xi$, to denote matrices whose matrix elements are given by $f_{ij}$, $C_l^{ij}$, and $\Xi_{ij}$, respectively. 

First of all, the GMV estimator can be made unbiased if we require
\begin{equation}
    1 = \int_{\vect L=\mathbf l_1-\mathbf l_2} f_{ij}(\vect l_1,\vect l_2)\,\Xi_{ij}(\vect l_1,\vect l_2).
\end{equation}
The variance of $\hat\alpha_{GMV}$ can be defined analogously to \refeq{HO_variance}. Under the condition of unbiasedness, the weights that minimize this variance are
\begin{equation}
    \vect \Xi(\vect l_1,\vect l_2) = \frac 12\,N_{GMV}(L)\,\vect C_{l_1}^{-1}\,\vect f(\vect l_1,\vect l_2)\,\vect C_{l_2}^{-1},
\label{eq:Xi}
\end{equation}
where the reconstruction noise spectrum $N_{GMV}(L)$ is given by the integral of a matrix trace
\begin{equation}
    [N_{GMV}(L)]^{-1} = \frac 12\,\int_{\vect L=\mathbf l_1-\mathbf l_2} {\rm Tr} \left[ \vect C_{l_1}^{-1}\,\vect f(\vect l_1,\vect l_2)\,\vect C_{l_2}^{-1}\,\vect f(\vect l_2,\vect l_1) \right].
\label{eq:noise_GMV}
\end{equation}
Using the symmetry $f_{YX}(\vect l_1,\vect l_2) = f_{XY}(\vect l_2,\vect l_1)$, we can rewrite $\vect f(\vect l_2,\vect l_1) = \vect f(\vect l_1,\vect l_2)^T$.

The precise gap between the GMV estimator and the HO estimator is more manifest if we recast \refeq{HO}) into the form
\begin{equation}
    \hat\alpha_{HO}(\vect L) = \sum_{i,j}\,\int_{\vect L=\mathbf l_1-\mathbf l_2} X^i(\vect l_1)\,X^{j*}(\vect l_2)\,w_{ij}(|\vect l_1-\vect l_2|)\,F_{ij}(\vect l_1,\vect l_2).
\end{equation}
A comparison with \refeq{GMV} suggests that $\hat\alpha_{HO}$ would be made to match $\hat\alpha_{GMV}$ precisely if we allowed $w_{ij}$ to depend {\it arbitrarily} on $\vect l_1$ and $\vect l_2$, i.e. $w_{ij}=w_{ij}(\vect l_1, \vect l_2)$, rather than on the combination $L = |\vect l_1-\vect l_2|$. In that case, we would have simply treated the product of $w_{ij}(\vect l_1, \vect l_2)$ and $F_{ij}(\vect l_1, \vect l_2)$ as a single coefficient $\Xi_{ij}(\vect l_1, \vect l_2)$, and hence followed the logic of constructing the GMV estimator by minimizing the variance.

Nevertheless, the conceptual distinction between the GMV and HO estimators does not lead to a drastic difference in the estimation error, as we demonstrate both analytically and numerically in \refsec{computation}.

\section{Maximum likelihood estimator}\label{sec:max_likelihood}

In this section, we sketch another approach to estimating birefringence rotation in the CMB observables from the perspective of maximizing the likelihood function. This approach is inspired by the work of Ref.~\cite{Hirata_2003weaklensing} on constructing the maximum likelihood estimator for weak gravitational lensing. By modifying the maximum likelihood estimator, a quadratic estimator can be derived, which we show is precisely the GMV estimator. The connection between the maximum likelihood estimator and the quadratic estimator in terms of Fisher matrices in the context of weak gravitational lensing has been noted in Ref.~\cite{Hirata_2003}.

Let $\tilde{\vect d}$ denote a column vector that collectively contains all the available CMB observables, $\tilde T(\vect l),\tilde E(\vect l),\tilde B(\vect l)$, for all wave vectors $\vect l$, free of polarization rotation or noise. The vector $\tilde{\vect d}$ has a covariance matrix:
\begin{equation}
    \tilde{\vect C} = \ev{\tilde{\vect d}\,\tilde{\vect d}^T},
\label{eq:cov_d}
\end{equation}
where random realizations of CMB anisotropies are averaged over. The covariance matrix $\tilde{\vect C}$ in principle follows from the theoretical prediction that the CMB anisotropies of cosmological origin are Gaussian random variables. 

While the Fourier components $\tilde T(\vect l)$, $\tilde E(\vect l)$ and $\tilde B(\vect l)$ are complex-valued, in principle a set of real-valued variables can be instead used to build the vector $\tilde{\vect d}$. Correspondingly, the covariance matrix is real symmetric and positive definite. For this reason, we shall have in mind the machinery of linear algebra for real variables, and apply it to the following derivations.

We note that there are no primary $B$ modes in the absence of primordial tensor perturbations; \cite{PhysRevLett.78.2054,Zaldarriaga_1997,PhysRevLett.78.2058} the cosmic $B$-mode polarization detected in the CMB results from weak lensing by the large-scale structure (see Ref.~\cite{LEWIS_2006} for a review), and for this reason the CMB anisotropies acquire mode couplings. Nevertheless, for our purpose it suffices to treat the $B$ modes in the likelihood function as if they were genuinely Gaussian random primary anisotropies, as long as the corresponding (isotropic) power spectrum is used in constructing $\tilde{\vect C}$.

The log likelihood of the vector $\tilde{\vect d}$ given the covariance matrix $\tilde{\vect C}$ is
\begin{equation}
    \ln\mathcal L[\tilde{\vect d}|\tilde{\vect C}] = -\frac 12\,\tilde{\vect d}^T\,\tilde{\vect C}^{-1}\,\tilde{\vect d} - \frac 12\,\ln\det\tilde{\vect C} + \mathrm{const}.
\label{eq:logL_data_only}
\end{equation}
In the presence of rotation $\alpha$ and noise $\vect n$, the measured CMB observables can be written as
\begin{equation}
    \vect d = \left[ \vect M(\alpha) \right]^{-1}\,\tilde{\vect d} + \vect n,
\end{equation}
where $\vect M(\alpha)$ is an orthogonal matrix (satisfying $[\vect M(\alpha)]^{-1} = [\vect M(\alpha)]^T$) that describes the action of pointwise polarization rotation on the anisotropies. Assuming that the noise $\vect n$ has a covariance matrix $\vect C_n$, the joint log likelihood of $\vect d$ and $\vect n$ is now
\begin{equation}
    \ln\mathcal L[\vect d,\vect n|\tilde{\vect C},\alpha,\vect C_n] = -\frac 12\,\vect n^T\,\vect C_n^{-1}\,\vect n - \frac 12\,(\vect d-\vect n)^T\,[\vect M(\alpha)]^T\,\vect{\tilde C}^{-1}\,\vect M(\alpha)\,(\vect d-\vect n),
\label{eq:joint_logL}
\end{equation}
where we have dropped irrelevant additive constant terms. For a fixed $\vect d$, \refeq{joint_logL} is maximized at
\begin{equation}
    \vect n = \vect n_{\mathrm{max}} = \left( \vect C_n^{-1} + \vect M^T\,\vect{\tilde C}^{-1}\,\vect M \right)^{-1}\,\vect M^T\,\vect{\tilde C}\,\vect M\,\vect d,
\end{equation}
and the maximum value of the likelihood is
\begin{equation}
    \ln\mathcal L[\vect d|\tilde{\vect C},\alpha,\vect n=\vect n_{\mathrm{max}}] = -\frac 12\,\vect d^T\,(\vect C_n + \vect M^T\,\tilde{\vect C}\,\vect M)^{-1}\,\vect d,
\label{eq:logL}
\end{equation}
which we shall simply call $\ln\mathcal L$. We can now estimate the rotation field by maximising this log likelihood with respect to $\alpha$. Anticipating $\alpha$ to be numerically small in radians \cite{Jain_2021}, we consider an approximation of $\ln\mathcal L$ as a second-order polynomial in $\alpha$:
\begin{equation}
    \ln\mathcal L = \mathrm{const} + \vect d^T\,\vect K_1(\alpha)\,\vect d - \frac 12\,\vect d^T\,\vect K_2(\alpha)\,\vect d + O(\alpha^3).
\end{equation}
The matrices $\vect K_1(\alpha)$ and $\vect K_2(\alpha)$ have linear and quadratic dependence on $\alpha$, respectively. To find these matrices, we first Taylor-expand $\vect M(\alpha)$:
\begin{equation}
    \vect M(\alpha) = \vect I - \vect T(\alpha) + O(\alpha^2),
\label{eq:taylor}
\end{equation}
where $\vect T(\alpha)$ is an {\it antisymmetric} matrix with linear dependence on $\alpha$. Substituting this expansion into \refeq{logL} and defining $\vect C = \tilde{\vect C} + \vect C_n$, we obtain
\begin{align}
    \vect K_1 &= \vect C^{-1}\,\vect T\,\tilde{\vect C}\,\vect C^{-1},\\
    \vect K_2 &= \vect C^{-1}\,\vect T^T\,\tilde{\vect C}\,\vect C^{-1}\,\tilde{\vect C}\,\vect T\,\vect C^{-1} + \vect C^{-1}\,\tilde{\vect C}\,\vect T^T\,\vect C^{-1}\,\vect T\,\tilde{\vect C}\,\vect C^{-1} - 2\,\vect C^{-1}\,\vect T^T\,\tilde{\vect C}\,\vect C^{-1}\,\vect T\,\tilde{\vect C}\,\vect C^{-1} \nonumber\\
    &\quad + \vect C^{-1}\,\vect T^T\,\vect T\,\tilde{\vect C}\,\vect C^{-1} - \vect C^{-1}\,\vect T^T\,\tilde{\vect C}\,\vect T\,\vect C^{-1}.
\end{align}
We can make the dependence on $\alpha$ explicit by writing
\begin{align}
    \vect T(\alpha) &= \sum_{\vect L}\,\vect T_{\vect L}\,\alpha_{\vect L},\label{eq:T_alpha}\\
    \vect d^T\,\vect K_1(\alpha)\,\vect d &= \sum_{\vect L}\,\mathcal K_{1,\vect L}\,\alpha_{\vect L},\\
    \vect d^T\,\vect K_2(\alpha)\,\vect d &= \sum_{\vect L}\,\sum_{\vect L'}\, \mathcal K_{2,\vect L \vect L'}\,\alpha_{\vect L}^*\,\alpha_{\vect L'},
\end{align}
where the indices $\vect L$ and $\vect L'$ label the components of the anisotropic rotation field on the sky (i.e. its various Fourier modes $\alpha(\vect L)$). With the cubic and higher order terms neglected, \refeq{logL} takes a quadratic form:
\begin{equation}
    \ln\mathcal L = \mathrm{const} + \sum_{\vect L}\,\mathcal K_{1,\vect L}\,\alpha_{\vect L} - \frac 12\,\sum_{\vect L}\,\sum_{\vect L'}\,\mathcal K_{2,\vect L \vect L'}\,\alpha_{\vect L}^*\,\alpha_{\vect L'}.
\end{equation}
This log likelihood is maximized for
\begin{equation}
    \alpha_{\vect L} = \sum_{\vect L'}\, [\mathcal K_2^{-1}]_{\vect L \vect L'}\,\mathcal K_{1,\vect L'},\label{eq:alpha_maximized}
\end{equation}
where the matrix inverse of $\mathcal K_2$ is taken with respect to the indices $\vect L$ and $\vect L'$. Note that the matrices $\mathcal K_{1,\vect L}$ and $\mathcal K_{2,\vect L \vect L'}$ both depend quadratically on the specific realization of the measured CMB observables $\vect d$.

\section{Relation between various quadratic estimators}\label{sec:relation}

\subsection{Equivalence to GMV estimator}

To make contact with quadratic estimators, we consider replacing the matrix $\mathcal K_2$ with $\bar{\mathcal K}_2 = \ev{\mathcal K_2}$, which is the ensemble expectation after averaging over random realizations of $\vect d$. The matrix $\mathcal K_2$ that correspond to any specific realization of $\vect d$ fluctuates around $\bar{\mathcal K}_2$, but the fluctuation is usually small in the limit of a large number of Fourier modes. Without having any statistically preferred direction on the sky, the matrix $\bar{\mathcal K}_2$ is shown below to be diagonal in $\vect L$ space. We therefore define the reconstructed rotation field as follows, by modifying \refeq{alpha_maximized}:
\begin{equation}
    \alpha_{\mathrm{rec}, {\vect L}} = \sum_{\vect L'}\, [\bar{\mathcal K}_2^{-1}]_{\vect L \vect L'}\,\mathcal K_{1, \vect L'}.
\end{equation}
Since the only dependence on the data $\vect d$ is in $\mathcal K_{1,\vect L'}$ as a bilinear form in $\vect d$, $\alpha_{\mathrm{rec},\vect L}$ is a quadratic estimator. After some algebra and using \refeq{cov_d}, we derive
\begin{equation}
    \bar{\mathcal K}_{2, \vect L \vect L'} = {\rm Tr}\left[\tilde{\vect C}\,\vect C^{-1}\,\tilde{\vect C}\,\vect T^T_{-\vect L}\,\vect C^{-1}\,\vect T_{\vect L'}\right] - {\rm Tr}\left[\tilde{\vect C}\,\vect C^{-1}\,\vect T^T_{-\vect L}\,\tilde{\vect C}\,\vect C^{-1}\,\vect T_{\vect L'}\right].
\label{eq:bar_K2}
\end{equation}

Finally, we show that $\alpha_{\mathrm{rec}, {\vect L}}$ coincides with the GMV estimator $\hat\alpha_{GMV}(L)$ constructed in \refsec{GMV}. To this end, we first recast some of the equations in the preceding sections in terms of matrices. \refeq{unrotated_power_spectra} becomes \refeq{cov_d}, while \refeq{linear_response} is replaced by
\begin{equation}
    \ev{[\vect M(\alpha)]^T\,\tilde{\vect d}\,\tilde{\vect d}^T\,\vect M(\alpha)} = \ev{\tilde{\vect d}\,\tilde{\vect d}^T} + \sum_{\vect L}\,\vect f_{\vect L}\,\alpha_{\vect L} + O(\alpha^2).
\end{equation}
In other words, an arbitrary matrix element of $\vect f_{\vect L}$ has the form $f_{ij}(\vect l_1,\vect l_2)\,(2\pi)^2\,\delta_D(\vect l_1-\vect l_2-\vect L)$. Using \refeq{taylor} and \refeq{T_alpha}, we find
\begin{equation}
    \vect f_{\vect L} = \vect T_{\vect L}\,\tilde{\vect C} + \tilde{\vect C}\,\vect T^T_{\vect L}.
\end{equation}
On the other hand, an arbitrary element of $\vect C$, denoted $C^{i\vect l_1,j\vect l_2}$, is equal to $C^{ij}_{l_1}\,(2\pi)^2\,\delta_D(\vect l_1-\vect l_2)$. With these in mind, we can rewrite \refeq{noise_GMV} as
\begin{align}
    (2\pi)^2\delta_D(\vect L-\vect L')N_{GMV}(L)^{-1} &= \frac 12\,{\rm Tr} \left[\vect C^{-1}\,\vect f_{-\vect L}\,\vect C^{-1}\,\vect f_{\vect L'} \right]\nonumber\\
    &= \frac 12\,{\rm Tr}\lf[\vect C^{-1}\,\vect T_{-\vect L}\,\tilde{\vect C}\,\vect C^{-1}\,\vect T_{\vect L'}\,\tilde{\vect C} + \vect C^{-1}\,\tilde{\vect C}\,\vect T^T_{-\vect L}\,\vect C^{-1}\,\vect T_{\vect L'}\,\tilde{\vect C}\ri.\nonumber\\
    &\quad + \lf.\vect C^{-1}\,\tilde{\vect C}\,\vect T^T_{-\vect L}\,\vect C^{-1}\,\vect T_{\vect L'}\,\tilde{\vect C} + \vect C^{-1}\,\tilde{\vect C}\,\vect T^T_{-\vect L}\,\vect C^{-1}\,\tilde{\vect C}\,\vect T^T_{\vect L'}\ri]\nonumber\\
    &= \bar{\mathcal K}_{2,\vect L \vect L'},\label{eq:N_GMV_barK2}
\end{align}
where in the last line we have used the cyclic property of the trace and the antisymmetry of $\vect T_L$ to match \refeq{bar_K2}. We note that $\bar{\mathcal K}_{2,\vect L\vect L'}$ is indeed diagonal in $\vect L$ space, as expected from statistical isotropy. Finally, \refeq{GMV} and \refeq{Xi} can be rewritten as
\begin{align}
    \hat\alpha_{GMV}(\vect L) &= \frac 12\,N_{GMV}(L)^{-1}\,\vect d^T\,\vect C^{-1}\,\vect f_{\vect L}\,\vect C^{-1}\,\vect d\\
    &= \frac 12\,\sum_{\vect L'}\,\bar{\mathcal K}_{2,\vect L \vect L'}\,\vect d^T\,\vect C^{-1}\,\left( \vect T_{\vect L'}\,\tilde{\vect C} + \tilde{\vect C}\,\vect T^T_{\vect L'} \right)\,\vect C^{-1}\,\vect d\\
    &= \sum_{\vect L'}\,\bar{\mathcal K}_{2,\vect L \vect L'}\,\vect d^T\,\vect C^{-1}\,\vect T_{\vect L'}\,\tilde{\vect C}\,\vect C^{-1}\,\vect d\\
    &= \sum_{\vect L'}\,\bar{\mathcal K}_{2,\vect L \vect L'}\,\mathcal K_{1\vect L'} = \hat\alpha_{\mathrm{rec}, \vect L}.
\end{align}

We note that the above derivation of the connection between the maximum likelihood estimator and GMV estimator assumes an antisymmetric $\vect T$ matrix (or more generally an anti-Hermitian matrix). In more familiar notation, if
\begin{equation}
    \begin{pmatrix}
    \delta T(\vect l)\\ \delta E(\vect l)\\ \delta B(\vect l)
    \end{pmatrix} = \int\frac{\rmd^2\vect l'}{(2\pi)^2} \vect W(\vect l,\vect l') \begin{pmatrix}
    \tilde T(\vect l')\\ \tilde E(\vect l')\\ \tilde B(\vect l')
    \end{pmatrix},
\end{equation}
for some $3\times 3$ matrix $\vect W$ that depends on $\vect l$ and $\vect l'$, then we require $\vect W^T(\vect l',\vect l) = -\vect W(\vect l,\vect l')$ [or more generally $\vect W^T(\vect l',\vect l) = -\vect W^*(\vect l,\vect l')$]. While this is true for polarization rotations, it is not true for gravitational lensing (see Eq.~(4) of \cite{Hu_2002}).

\subsection{Equivalence between HO and GMV estimators}

Having summarized the quadratic estimators, we now quantify their performance using the power spectrum of reconstruction noise, namely the variance of $\hat\alpha(\vect L)$ as a function of $L=|\vect L|$. We will perform calculations under the null hypothesis as compelling evidence for anisotropic CMB birefringence has not yet been found. Since we have already shown in \refsec{max_likelihood} that the simple modification of the maximum likelihood estimator is equivalent to the GMV estimator, it remains to compare the GMV and HO estimators. We show that, if the two are constructed using only the polarization but not the temperature anisotropies, then they have the same reconstruction noise power spectrum. When the temperature anisotropies are included in estimating rotation, we find numerically in \refsec{computation} that the difference in the reconstruction noise spectra between the GMV and HO estimators is insignificant.


From \refeq{F_TE} and \refeq{lambda_TE}, we can see that the presence of a non-vanishing cross-variable spectrum, $C^{TE}_l$, makes formulas more complicated (note that $C^{EB}_l$ vanishes due to statistical invariance under parity). This complication is avoided if temperature anisotropies $T(\vect l)$ are entirely excluded from the construction of the GMV and HO estimators. When only the $E$ and $B$ polarization anisotropies are considered, it is possible to show that the GMV and HO estimators are equivalent. In fact, this is a special case of the general statement that the GMV and HO estimators are identical whenever $C^{XY}_l=0$ for any $X\neq Y$, which we now show.

Assuming $C^{XY}_l=0$ for $X\neq Y$, \refeq{covariance} implies that $\vect N$ is a diagonal matrix, \ie there is no covariance between different single-pair estimators. Using \refeq{alpha_XY}, \refeq{F_XY}, and \refeq{HO}, the HO estimator becomes
\begin{equation}
    \hat\alpha_{HO}(\vect L) = N_{HO}(L) \sum_{(X, Y)}\int_{\vect L=\vect l_1-\vect l_2} X_o(\vect l_1)\,Y^*_o(\vect l_2)\,\frac{f_{XY}(\vect l_1,\vect l_2)}{(1+\delta_{XY})\,C^{XX}_{l_1}\,C^{YY}_{l_2}},
\end{equation}
where the sum is over unique pairs of observables $XY$, and
\begin{equation}
    N_{HO}(L) = \lf(\sum_{(X, Y)} N^{-1}_{XY}(L)\ri)^{-1}.
\end{equation}
On the other hand, \refeq{GMV} can be rewritten as
\begin{equation}
    \hat\alpha_{GMV} = N_{GMV}(L) \sum_{(X, Y)}\int_{\vect L=\vect l_1-\vect l_2} X_o(\vect l_1)\,Y^*_o(\vect l_2)\,\frac{f_{XY}(\vect l_1,\vect l_2)}{(1+\delta_{XY})\,C^{XX}_{l_1}\,C^{YY}_{l_2}}.
\end{equation}
The equality of the reconstruction noise spectra $N_{HO}(L)=N_{GMV}(L)$ can be straightforwardly derived from \refeq{noise_GMV} and the symmetry property of $\vect f(\vect l_1,\vect l_2)$. Therefore, we conclude that $\hat \alpha_{HO}(L) = \hat \alpha_{GMV}(L)$ if $C^{XY}_l=0$ for $X\neq Y$.

In the special case where only polarization anisotropies $E$ and $B$ are considered, the reconstruction noise spectrum has the form
\begin{equation}
    N_{HO}(L) = N_{GMV}(L) = \lf(\int_{\vect L=\mathbf l_1-\mathbf l_2} \lf[\frac{f_{EE}(\vect l_1,\vect l_2)^2}{2\,C_{l_1}^{EE}C_{l_2}^{EE}} + \frac{f_{EB}(\vect l_1,\vect l_2)^2}{C_{l_1}^{EE}C_{l_2}^{BB}} + \frac{f_{BB}(\vect l_1,\vect l_2)^2}{2\,C_{l_1}^{BB}C_{l_2}^{BB}}\ri]\ri)^{-1}.
\end{equation}

\section{Simultaneous estimation of lensing and rotation}
\label{sec:simultaneous}

We have carried out the analysis under the assumption that the power spectra of lensed (but unrotated) CMB observables are known. The estimator for the rotation $\hat\alpha$ is constructed from these lensed power spectra through the weights $F_{XY}(\vect l_1,\vect l_2)$ or $\Xi_{ij}(\vect l_1,\vect l_2)$. However, one might worry that the estimate for lensing is contaminated by the presence of anisotropic rotation, and conversely the estimate for rotation might be contaminated by lensing. We show in this section that any dependence of quadratic estimators $\ev{\hat\alpha_{XY}}$ on the lensing potential $\phi$ is of $\mathcal{O}(\phi^2)$, and likewise any dependence of quadratic estimators for the lensing potential $\ev{\hat\phi_{XY}}$ on the polarization rotation is of order $\mathcal{O}(\alpha^2)$. In other words, at linear order $\alpha$ and $\phi$ can be simultaneously estimated from the CMB anisotropies without bias \cite{Gluscevic2009Derotate}. However, the second-order contamination of the rotation power spectrum due to lensing will eventually need to be corrected for in future experiments (see Fig.~2 in Ref.~\cite{Namikawa_2017}).

We first rewrite \refeq{linear_response} to include the linear response due to the lensing potential:
\begin{equation}
    \ev{X(\vect l_1)Y^*(\vect l_2)} = \ev{\bar X(\vect l_1)\bar Y^*(\vect l_2)} + f_{XY}^\phi(\vect l_1,\vect l_2)\phi(\vect l_1-\vect l_2) + f_{XY}^\alpha(\vect l_1,\vect l_2)\alpha(\vect l_1-\vect l_2) + O(\alpha^2,\phi^2,\alpha\phi),
\end{equation}
where $\bar X,\bar Y$ denote the unlensed {\it and} unrotated anisotropies. We now write down the ``naive'' minimum-variance estimator for $\alpha$ and $\phi$, respectively, from the observed quantities (with subscript $o$), pretending that only one effect takes place at a time:
\begin{align}
    \hat\alpha_{XY} &= \int_{\vect L=\vect l_1-\vect l_2}X_o(\vect l_1)Y_o^*(\vect l_2)F^\alpha_{XY}(\vect l_1,\vect l_2),\\
    \hat\phi_{XY} &= \int_{\vect L=\vect l_1-\vect l_2}X_o(\vect l_1)Y_o^*(\vect l_2)F^\phi_{XY}(\vect l_1,\vect l_2).
\end{align}
Here the weights $F^\alpha_{XY}$ are the same as in \refeq{F_XY} and \refeq{F_TE}. The weights $F^\phi_{XY}$ are defined in Ref.~\cite{Maniyar_2021}.

For $\vect L\neq 0$, the rotation estimator has an expectation value
\begin{align}
    \ev{\hat\alpha_{XY}(\vect L)} &= \int_{\vect L=\vect l_1-\vect l_2} \ev{\bar X(\vect l_1)\bar Y^*(\vect l_2)}\,F^\alpha_{XY}(\vect l_1,\vect l_2) + \int_{\vect L=\vect l_1-\vect l_2} \ev{\bar X_n(\vect l_1)\bar Y_n^*(\vect l_2)}\,F^\alpha_{XY}(\vect l_1,\vect l_2) \nonumber\\
    &\quad + \alpha(\vect L)\int_{\vect L=\vect l_1-\vect l_2} f^\alpha_{XY}(\vect l_1,\vect l_2)\,F^\alpha_{XY}(\vect l_1,\vect l_2) + \phi(\vect L)\int_{\vect L=\vect l_1-\vect l_2} f^\phi_{XY}(\vect l_1,\vect l_2)\,F^\alpha_{XY}(\vect l_1,\vect l_2),
\end{align}
where $f^\alpha_{XY}(\vect l_1, \vect l_2)$ are found in \reftab{response_funcs}, and $f^\phi_{XY}$ are found in Table 1 of Ref.~\cite{Maniyar_2021}. The first two terms vanish because $\vect l_1\neq \vect l_2$. The third term reduces to $\alpha(\vect L)$ due to the unbiasedness of the estimator (\refeq{unbiased}). It remains to see whether the last term vanishes, or in other words, whether the naive estimator for $\alpha$ is contaminated by lensing.

Here, we show explicitly the proof for $XY = EB$, but the result holds analogously for all pairs $XY$.
\begin{align}
    &\quad \int_{\vect L=\vect l_1-\vect l_2} f^\phi_{EB}(\vect l_1,\vect l_2)\,F^\alpha_{EB}(\vect l_1,\vect l_2)\\
    &= \int_{\vect L=\vect l_1-\vect l_2} \frac{f^\phi_{EB}(\vect l_1,\vect l_2)\,f^\alpha_{XY}(\vect l_1,\vect l_2)}{C^{EB}_{l_1}C^{EB}_{l_2}}\\
    &= \int\frac{\rmd^2\vect l_1\,\rmd^2\vect l_2}{(2\pi)^2}\delta_D(l_{1x}-l_{2x}-L_x)\,\delta_D(l_{1y}-l_{2y}-L_y)\,\left(C^{EB}_{l_1}\,C^{EB}_{l_2}\right)^{-1}\\
    &\quad \times \lf(\tilde C^{E\nabla E}_{l_1}\,\vect L\cdot\vect l_1 + \tilde C^{B\nabla B}_{l_2}\,\vect L\cdot\vect l_2\ri)\,\sin 2\varphi_{12}\times 2\,\lf(\tilde C^{BB}_{l_1} - \tilde C^{EE}_{l_2}\ri)\,\cos 2\varphi_{12}.
\end{align}
The quantities $\tilde C^{X\nabla Y}_l$ are the lensed gradient spectrum, defined in \cite{Lewis_2011,Fabbian_2019}. Under a change of variables: $l_{iy}\to -l_{iy}$ for $i=1,2$, we have $\varphi_{12} \to -\varphi_{12}$ and
\begin{align}
    \delta_D(l_{1y}-l_{2y}-L_y)\,\vect L\cdot\vect l_i &\to \delta_D(-l_{1y}+l_{2y}-L_y)\,(L_xl_{ix} - L_yl_{iy})\\
    &= \delta_D(-l_{1y}+l_{2y}-L_y)\,[L_xl_{ix} + (l_{1y}-l_{2y})\,l_{iy}]\\
    &= \delta_D(l_{1y}-l_{2y}-L_y)\,[L_xl_{ix} + (l_{1y}-l_{2y})\,l_{iy}]\\
    &= \delta_D(l_{1y}-l_{2y}-L_y)\,\vect L\cdot\vect l_i.
\end{align}
The whole integrand is odd under this change of variables, and hence the integral vanishes.

By putting together \reftab{response_funcs}, and Table 1 of Ref.~\cite{Maniyar_2021}, it is easy to see that the corresponding integrals for other pairs $XY$ also vanish. We conclude that
\begin{equation}
    \ev{\hat\alpha_{XY}(\vect L)} = \alpha(\vect L) + O(\alpha^2,\phi^2,\alpha\phi),
\end{equation}
where $\hat\alpha_{XY}$ here is constructed from unlensed and unrotated spectra. A similar statement can be made about $\hat\phi_{XY}(\vect L)$ by the same parity argument. Ref.~\cite{Gluscevic2009Derotate} reached the same conclusion based on the parity argument, for the full-sky version of these estimators.

\section{Numerical forecasts}
\label{sec:computation}

We compute the reconstruction noise spectra for the HO and GMV estimators using polarization and temperature anisotropies. The lensed CMB power spectra ($\tilde C^{XY}_l$) are generated by \texttt{CAMB} \cite{CAMB} up to $l_{\mathrm{max}}=3000$ \cite{Planck_2018_vi}, with concordance cosmological parameters $H_0=67.4\,\mathrm{km\,s^{-1}\,s^{-1}\,Mpc}$, $\Omega_b h^2=0.0224$, $\Omega_c h^2=0.120$, $m_\nu=0.06\,\mathrm{eV}$, $\Omega_K=0$, and $\tau=0.054$.

We assume that the noise is Gaussian, with an isotropic power spectrum in the form \cite{Hu_2002}
\begin{equation}
    C^{XY,n}_l = \delta_{XY}\,\Delta_X^2\,e^{(l^2\,\Theta^2)/(8\ln 2)},
\end{equation}
where $\Theta$ is the full width at half maximum (FWHM) of the Gaussian beam profile, and $\Delta_X=\Delta_T$ and $\Delta_P$ for temperature ($X=T$) and polarization ($X=E,\,B$), respectively. We assume a fully polarized detector, with $\sqrt 2\,\Delta_T = \Delta_P$~\cite{Hu_2002}.

We perform calculations for three sets of detector parameters $\Delta_T$ and $\Theta$, which are appropriate for a range of CMB experiments: Planck 2018 \cite{Planck_2018_iv}, Simons Observatory \cite{Ade_2019}, and a Stage-IV CMB experiment (CMB-S4) \cite{abazajian2016cmbs4}. See Ref.~\cite{Pogosian_2019} for details on the the detection of birefringence using the latter two experiments. In all three experiments, we include primary anisotropy modes in the range $l_{\rm min} \leqslant l \leqslant l_{\rm max}$. Refer to \reftab{params} for the parameter values we use.

\begin{table}[h!]
    \centering
    \begin{tabular}{|c|c|c|c|c|c|}
        \hline
        & $\Delta T\,[\mathrm{\mu K\,arcmin}]$ & $\Theta\, [\mathrm{arcmin}]$ & $l_{\mathrm{min}}$ & $l_{\mathrm{max}}$ & $f_{\mathrm{sky}}$\\
        \hline
        Planck SMICA & 45.0 & 5.0 & 5 & 3000 & 0.7\\
        Simons Observatory & 7.0 & 1.4 & 30 & 3000 & 0.4\\
        CMB-S4 & 1.0 & 1.4 & 30 & 3000 & 0.4\\
        \hline
    \end{tabular}
    \caption{Benchmark experimental parameters considered in this work \cite{Ferraro_2018}. }
    \label{tab:params}
\end{table}

\begin{figure}
    \begin{center}
        \scalebox{0.65}{\input{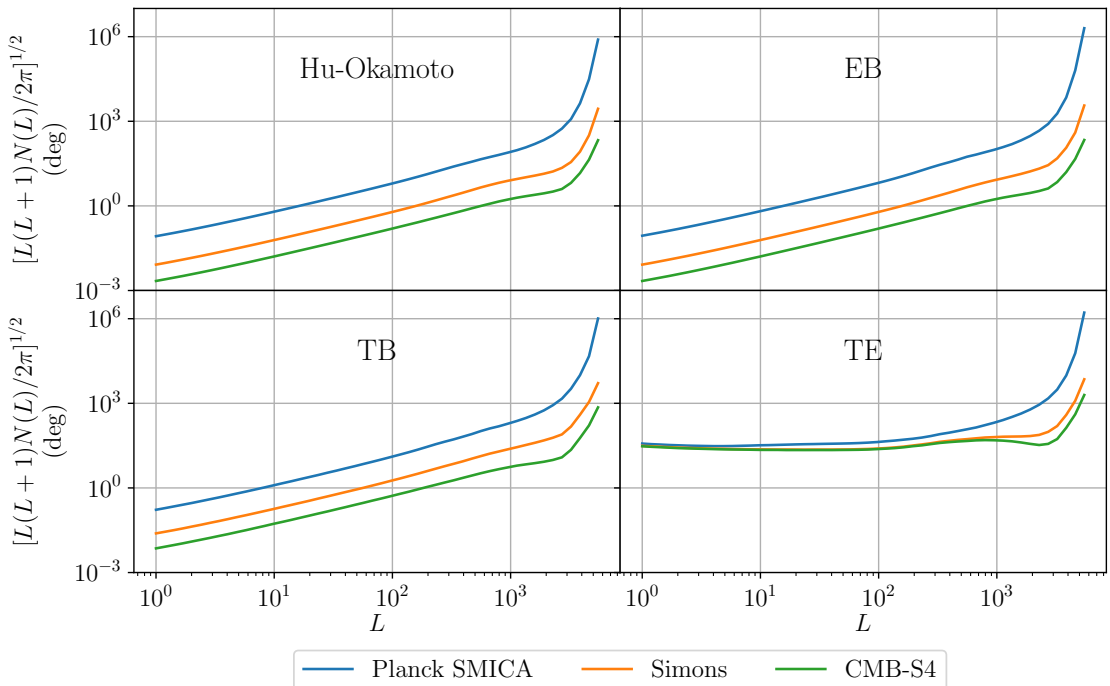}}
    \end{center}
    \caption{Reconstruction noise spectra for the HO estimator computed for the three sets of experimental parameters considered in this work. We plot the reconstruction noise spectra for the combined HO estimator, as well as those for using only one pair of CMB observables at a time with the lowest reconstruction noise. That is, we selectively present the results for three such pairs that make the greatest contribution to the overall detectability of the Hu-Okamoto estimator.}
    \label{fig:noise_curves}
\end{figure}

Mathematically, the reconstruction noise of the GMV estimator $N_{GMV}(L)$ is strictly smaller than or equal to that for the HO estimator $N_{HO}(L)$. Numerically, we find that the fractional difference between the two is no greater than 0.5\% for $L=1-10^3$. For this reason, we have not included separate curves for $N_{GMV}(L)$ in \reffig{noise_curves}, as the difference is too small to be visually discernible (they would overlap with the curves in the top-left plot for the Hu-Okamoto estimator).


The goal of estimating spatially dependent polarization rotation is to detect any new physics that might possibly cause this phenomenon, such as a pseudoscalar axion-like field that couples to the electromagnetic field. In this section, we study the constraining power of the quadratic estimators on the scenario that the rotation results from a network of cosmic strings. As explained in Ref.~\cite{Agrawal_2019CMBmillikan}, these are topological remnants reminiscent of a phase transition in the pseudoscalar field in the primordial Universe.

Since the cosmic strings that comprise the string network have random configurations, a key theoretical prediction is the power spectrum for the rotation field $\alpha$:
\begin{equation}
    \ev{\alpha(\vect L)\alpha(\vect L')} = (2\pi)^2\,\delta_D(\vect L-\vect L')\,K\,P(L),
\end{equation}
where the notation for ensemble averaging $\langle\cdots\rangle$ is over all string network realizations. Here $P(L)$ is a fiducial functional form for the power spectrum, and $K$ is a normalization constant that we would like to detect or constrain. In the presence of strings, a general quadratic estimator $\hat\alpha(\vect L)$ has a covariance
\begin{equation}
    \ev{\hat\alpha(\vect L)\,\hat\alpha(\vect L')} = (2\pi)^2\,\delta_D(\vect L-\vect L')\,\left( N(L) + K\,P(L) \right),
\end{equation}
where $N(L)$ is the null-hypothesis reconstruction noise power spectrum as we have discussed in the previous sections. In practical analysis, one works with a discretized set of wave vectors $\vect L$, and we may write
\begin{equation}
    \ev{\hat\alpha_{\vect L}\hat\alpha_{\vect L'}} = A\,\delta_{\vect L,\vect L'}\,\left( N_L + K\,P_L \right),
\end{equation}
where $A$ is the total area of the sky that is surveyed.

An unbiased estimator for the normalization $K$ may then be constructed from each $\hat\alpha(\vect L)$ by
\begin{equation}
    \hat K_{\vect L} = P_L^{-1} (A^{-1} |\hat\alpha_{\vect L}|^2 - N_L).\label{eq:K_def}
\end{equation}
To take full advantage of the information, these estimators can then be optimally combined to make a single estimator:
\begin{equation}
    \hat K = \sum_{\vect L}\,w_{\vect L}\,\hat K_{\vect L},
\end{equation}
The weights $w_{\vect L}$ are chosen to minimize the variance of $\hat K$ in the absence of polarization rotation (i.e. $K=0$):
\begin{equation}
    w_{\vect L} = \frac{\sum_{\vect L'} [\mathcal C_K^{-1}]_{\vect L,\vect L'}}{\sum_{\vect L',\vect L''} [\mathcal C_K^{-1}]_{\vect L',\vect L''}},
\end{equation}
where $\mathcal C_K$ is the covariance matrix of $K_{\vect L}$'s for the various $\vect L$'s. The variance of $\hat K$ is then
\begin{equation}
    \langle\hat K^2\rangle = \left[ \sum_{\vect L,\vect L'} [\mathcal C_K^{-1}]_{\vect L,\vect L'} \right]^{-1}.
\end{equation}
In fact, the covariance matrix is diagonal:
\begin{equation}
    (\mathcal C_K)_{\vect L,\vect L'} = \delta_{\vect L,\vect L'}\langle\hat K_{\vect L}^2\rangle,
\end{equation}
which is a consequence of statistical isotropy of the CMB anisotropies, namely that no direction on the sky has special statistics. While statistical isotropy holds for the primordial anisotropies, weak lensing by a specific realization of the foreground large-scale structure induces couplings between different Fourier modes \cite{Hu_2002}. For the purpose of estimating the signal-to-noise ratio of $\hat K$, we neglect the mode couplings due to lensing. This omission is justified, as we will show in \refsec{simultaneous} that any dependence of the rotation quadratic estimators on the lensing potential $\phi$ is of order $\mathcal O(\phi^2)$; when random realizations of foreground lens are averaged over, statistical isotropy is restored.

After some calculation, we have
\begin{equation}
    \langle \hat K_{\vect L}^2\rangle = \lf(\frac{N_L}{P_L}\ri)^2.
\end{equation}
The variance of $\hat K$ under the null hypothesis is given by
\begin{equation}
    \langle\hat K^2\rangle^{-1} = \sum_{\vect L} \lf(\frac{P_L}{N_L}\ri)^2.
\end{equation}
The square of the signal-to-noise ratio is
\begin{equation}
    \frac{K^2}{\langle\hat K^2\rangle} = \sum_{\vect L} \lf(\frac{KP_L}{N_L}\ri)^2 = \sum_L f_{\mathrm{sky}}(2L+1)\lf(\frac{KP_L}{N_L}\ri)^2, \label{eq:snr}
\end{equation}
where $f_{\mathrm{sky}}$ is the fraction of the full sky that is surveyed, and $2L+1$ is the number of discrete $\vect L$ vectors at a fixed magnitude $L$.

\section{String network models}
\label{sec:models}

As an example, we consider phenomenological models of string networks developed in Ref.~\cite{Jain_2021}. The theoretical framework is as follows. The Universe is populated by many circular string loops oriented in random directions, arising from a pseudoscalar field with anomaly coefficient $\mathcal A$. At any given conformal time $\tau$, the number of string loops with radius between $r$ and $r+\mathrm dr$ is given by the distribution function $\nu(r,\tau)\,\mathrm d r$, specified by the model.

In order to compare models with different distributions $\nu$, a dimensionless quantity is defined:
\begin{equation}
    \xi_\infty(\tau) = \frac{1}{a(\tau)^2 H(\tau)^2} \int_0^\infty \mathrm dr \, 2\pi r\,\nu(r,\tau).
\end{equation}
Intuitively, if all the string loops at conformal time $\tau$ were rearranged into a minimal number of Hubble-scale loops while conserving the energy density of the string network, then $\xi_\infty(\tau)$ would be equal to the number of such Hubble-scale loops per Hubble volume. Equivalently, $\xi_\infty(\tau)$ is the total length of strings in units of the Hubble scale per Hubble volume at conformal time $\tau$. Both models considered here are \emph{in scaling}. That is to say, the string network's energy density scales like the dominant energy density in the Universe, and $\xi_\infty(\tau) \equiv \xi_0$ is independent of $\tau$.\footnote{Whether cosmic string networks exhibit this in-scaling property is a matter of debate, with some studies concluding the affirmative and others finding a logarithmic growth in $\xi_\infty$. See Ref.~\cite{Jain_2021} for a brief discussion and references on this subject.} The model parameter $\xi_0$ is also referred to as the effective number of strings per Hubble volume.

The differences and additional parameters of these models are briefly stated below.
\begin{itemize}
    \item \textbf{Model I}: All string loops have the same radius $\zeta_0$ (in units of the Hubble scale at each redshift).
    
    \item \textbf{Model II}: A fraction $f_{\mathrm{sub}}$ of the string loops are at the sub-Hubble scale, with radii following a continuous, logarithmically flat distribution between $\zeta_{\mathrm{min}}=0.1$ and $\zeta_{\mathrm{max}}=1$. The remaining $(1-f_{\mathrm{sub}})$ fraction of the string loops have the same radius $\zeta_{\mathrm{max}}=1$.
    
\end{itemize}

For both models, the predicted power spectrum of the rotation field due to the string network is proportional to the product $\mathcal A^2\,\xi_0$:
\begin{equation}
    \ev{\alpha_{\vect L}\,\alpha_{\vect L'}} = A\,\delta_{\vect L,\vect L'}\,\mathcal A^2\,\xi_0\, P_L(\cdots),
\end{equation}
where the fiducial power spectrum shape $P_L(\cdots)$ does not depend on either $\mathcal A$ or $\xi_0$, but only on $\zeta_0$ for Model I or $f_{\mathrm{sub}}$ for Model II. The formalism we have laid down above is therefore applicable to constraining the overall normalization parameter $K=\mathcal A^2\,\xi_0$, given the value of the remaining model parameter.

\begin{figure}
    \begin{center}
        \scalebox{0.6}{\input{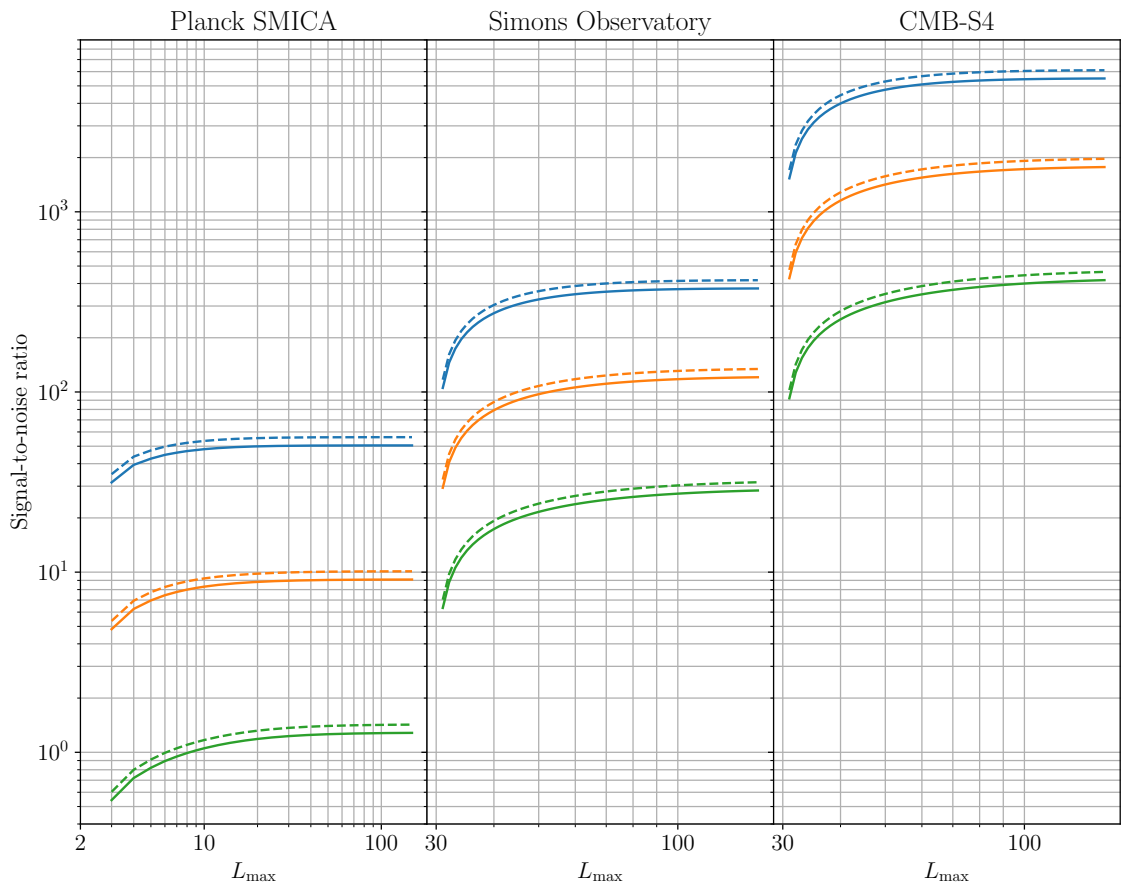}}
    \end{center}
    \caption{Signal-to-noise ratio (SNR) for detecting the polarization rotation power spectrum normalization $K=\mathcal A^2\,\xi_0$, as a function of $L_{\mathrm{max}}$ for string network Model I. The solid curves correspond to $\mathcal A=1$ and $\xi_0=10$, and the dashed curves correspond to $\mathcal A=1/3$ and $\xi_0=100$. The blue, orange, and green curves are computed for three different string loop radii $\zeta_0=1,10^{-0.5},10^{-1}$, respectively. The SNR saturates at $L_{\rm max} \gtrsim 30$ for Planck SMICA with $L_{\mathrm{min}}=2$. For the ground-based Simons Observatory and CMB-S4 with $L_{\mathrm{min}}=30$, the SNR saturates at $L_{\rm max} \gtrsim 100$.}
    \label{fig:snr_model1}
\end{figure}

\begin{figure}
    \begin{center}
        \scalebox{0.6}{\input{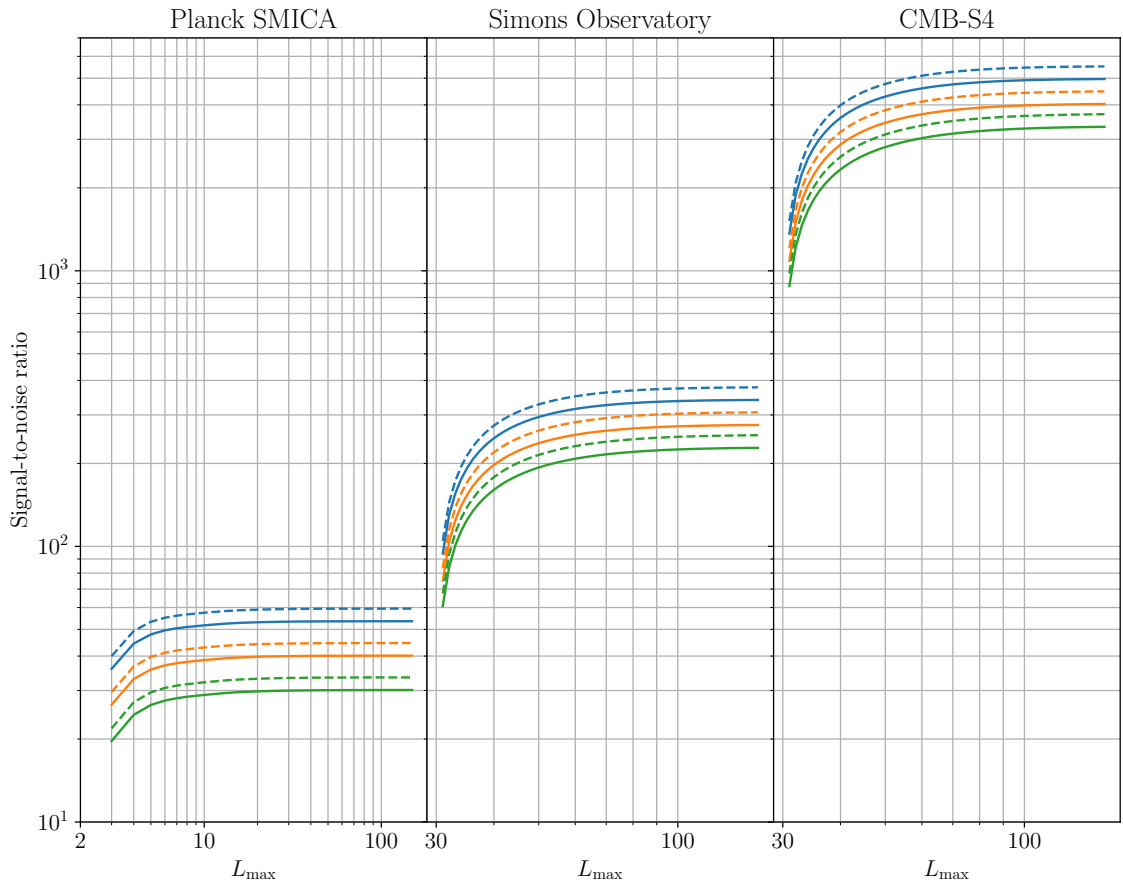}}
    \end{center}
    \caption{Signal-to-noise ratio (SNR) for detecting the polarization rotation power spectrum normalization $K=\mathcal A^2\,\xi_0$, as a function of $L_{\mathrm{max}}$ for string network Model II. Assuming $\zeta_{\mathrm{min}}=0.1$ and $\zeta_{\mathrm{max}}=1$, the blue, orange, and green curves are computed for three different fractions of sub-Hubble scale strings $f_{\mathrm{sub}}=0.2,0.6,0.9$, respectively. The SNR saturates at $L_{\rm max} \gtrsim 20$ for Planck SMICA with $L_{\mathrm{min}}=2$. For the ground-based Simons Observatory and CMB-S4 with $L_{\mathrm{min}}=30$, the SNR saturates at $L_{\rm max} \gtrsim 100$.}
    \label{fig:snr_model2}
\end{figure}

The overall SNR as in \refeq{snr} accumulates as we include more discrete $\vect L$ modes. In \reffig{snr_model1} and \reffig{snr_model2}, we plot the SNR of the $\hat K$ estimator as a function of $L_{\mathrm{max}}$, the maximum magnitude of $\vect L$ included into the analysis, for the three experimental setups we consider in \refsec{computation} and the two models stated above. For the large scale cutoff, we set $L_{\mathrm{min}}=2$ for Planck SMICA and $L_{\mathrm{min}}=30$ for Simons Observatory and CMB-S4. The survey sky coverage $f_{\mathrm{sky}}$ used for each experimental setup can be found in \reftab{params}. The estimator $\hat K$ is defined from the HO estimator through \refeq{K_def}, and there would be a negligible effect on the SNR curves if the GMV estimator were used instead of the HO estimator.

As shown in \reffig{snr_model1} and \reffig{snr_model2}, for all three experimental setups and both models considered, the SNR accumulates as $L_{\mathrm{max}}$ increases until it is saturated above $L_{\mathrm{max}}= 100$. This is a result of the string network producing most of the correlated rotations on very large angular scales \cite{Agrawal_2019CMBmillikan}. Therefore, it is sufficient to consider quadratic estimators $\hat\alpha(\vect L)$ for $|\vect L|<100$ for the purpose of constraining cosmic strings signals.

For the Planck SMICA experiment, we have chosen $L_{\mathrm{min}}=2$ on account of the analysis of Ref.~\cite{planck_constraint_2017}. An alternative choice of $L_{\mathrm{min}}=5$ results in the reduction of the SNR by a little less than a factor of two.

For the more simplistic model of uniform string sizes (Model I), \reffig{snr_model1} shows that the SNR depends strongly on the value of the string loop radius $\zeta_0$. As an improvement upon Model I, Model II exhibits a much weaker dependence of the SNR on the model parameter $f_{\mathrm{sub}}$.

According to \reffig{snr_model2}, we note that the Planck SMICA data should have already been sufficient for placing tight constraints on the more realistic string model with a continuous distribution of string loop sizes (Model II). In \refsec{bounds}, we obtain such a constraint using the rotation power spectrum estimated from the Planck 2015 data by Ref.~\cite{planck_constraint_2017}.

\begin{figure}
    \begin{center}
        \scalebox{0.7}{\input{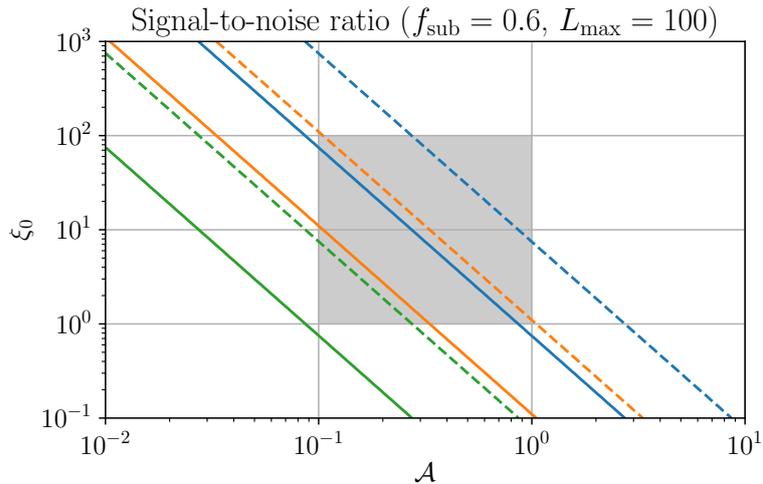}}
    \end{center}
    \caption{Signal-to-noise ratio of the $\hat K$ estimator in the $(\mathcal A,\xi_0)$ parameter space, assuming $f_{\mathrm{sub}} = 0.6$ and $L_{\mathrm{max}} = 100$. Solid lines mark the $3\sigma$ levels, while dashed lines mark the $30\sigma$ levels. The blue, orange, and green levels refer to the Planck SMICA, Simons Observatory, and CMB-S4 experiments, respectively. The theoretically favored region of the parameter space, $0.1 < \mathcal A < 1$ and $1 < \xi_0 < 100$, is indicated as the grey shaded area.}
    \label{fig:snr_param_space}
\end{figure}

Fixing $f_{\mathrm{sub}} = 0.6$ and $L_{\mathrm{max}} = 100$, we plot in \reffig{snr_param_space} the SNR contours in the $(\mathcal A,\xi_0)$ parameter plane.

\section{Planck 2015 bounds}
\label{sec:bounds}

In Ref.~\cite{planck_constraint_2017}, Planck 2015 data has been used to extract a (binned) birefringence power spectrum for $1\leq L\leq 730$. For both models studied in Ref.~\cite{Jain_2021} (stated in \refsec{models}), we obtain posteriors for the relevant model parameters using the Markov Chain Monte Carlo (MCMC) method, with particular emphasis on the power spectrum normalization $\mathcal A^2\,\xi_0$. Assuming $\mathcal A$ is an $\mathcal O(1)$ number, this allows us to place a constraint on the number of cosmic strings in the Universe.

The theoretical birefringence power spectra predicted by the two models are computed by direct integration according to Ref.~\cite{Jain_2021}. To compare with the model predictions, we only include the birefringence power spectrum for $2\leq L\leq 305$, where the bin sizes are $\Delta L=1$ for $2\leq L\leq 30$ and $\Delta L=50$ for $31\leq L\leq 330$. The dipole mode ($L=1$) has been excluded from this comparison, as it is most susceptible to contamination by residual foreground that has not been accounted for \cite{planck_constraint_2017}. The high-$L$ modes have also been excluded, as we anticipate little to no constraining power from their inclusion (see \reffig{snr_model1} and \reffig{snr_model2}).

\begin{figure}
    \centering
    \includegraphics{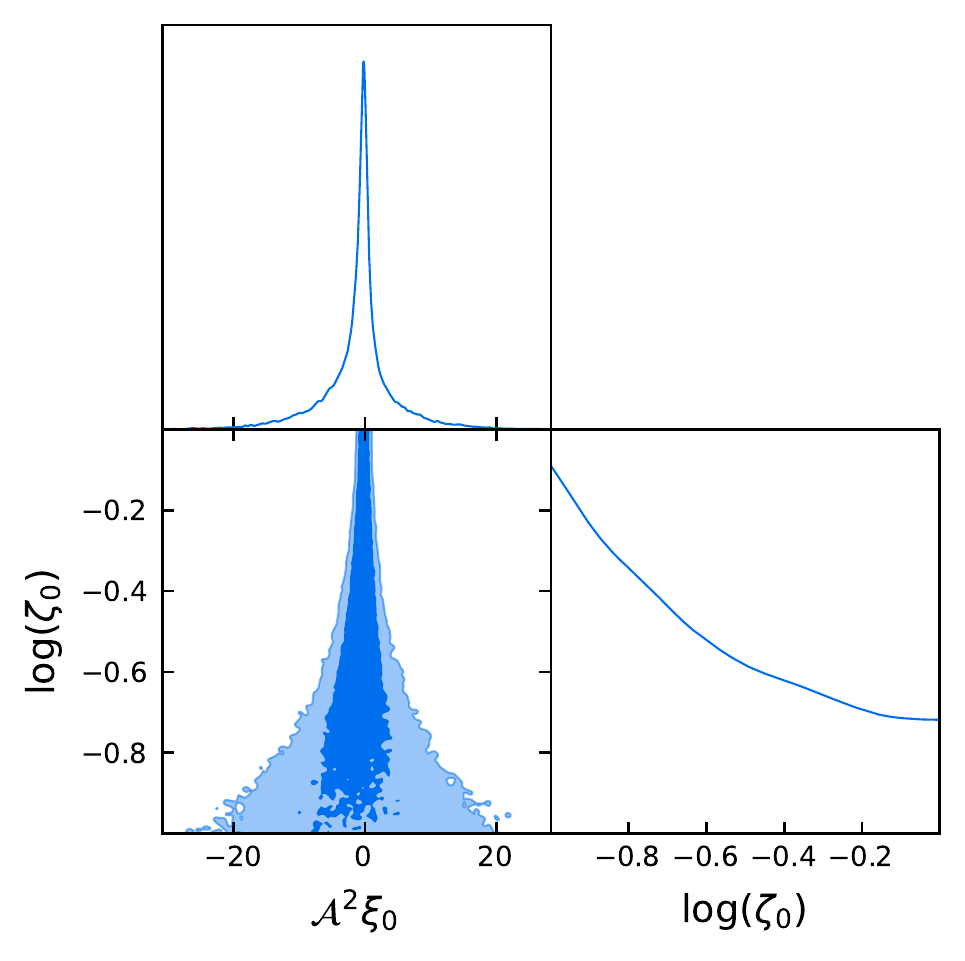}
    \caption{2D joint distribution and 1D marginalized distributions for the power spectrum normalization $\mathcal A^2\,\xi_0$ and the string loop radius $\zeta_0$ for string network Model I using Planck 2015 data from Ref.~\cite{planck_constraint_2017}.}
    \label{fig:triangle_model1}
\end{figure}

We present in \reffig{triangle_model1} the 2D joint distribution and 1D marginalized distributions for the power spectrum normalization $\mathcal A^2\,\xi_0$ and string loop radius $\zeta_0$ for Model I, obtained through MCMC sampling. Although the marginalized distribution of $\mathcal A^2\,\xi_0$ is highly peaked near 0, it also has a wide tail, leading to poorer constraining power of this model. Setting a flat prior in $\zeta_0$ for $0.1<\zeta_0<1$, we report an upper bound of $\mathcal A^2\,\xi_0<8.0$ at the 95\% confidence level.

\begin{figure}
    \centering
    \includegraphics{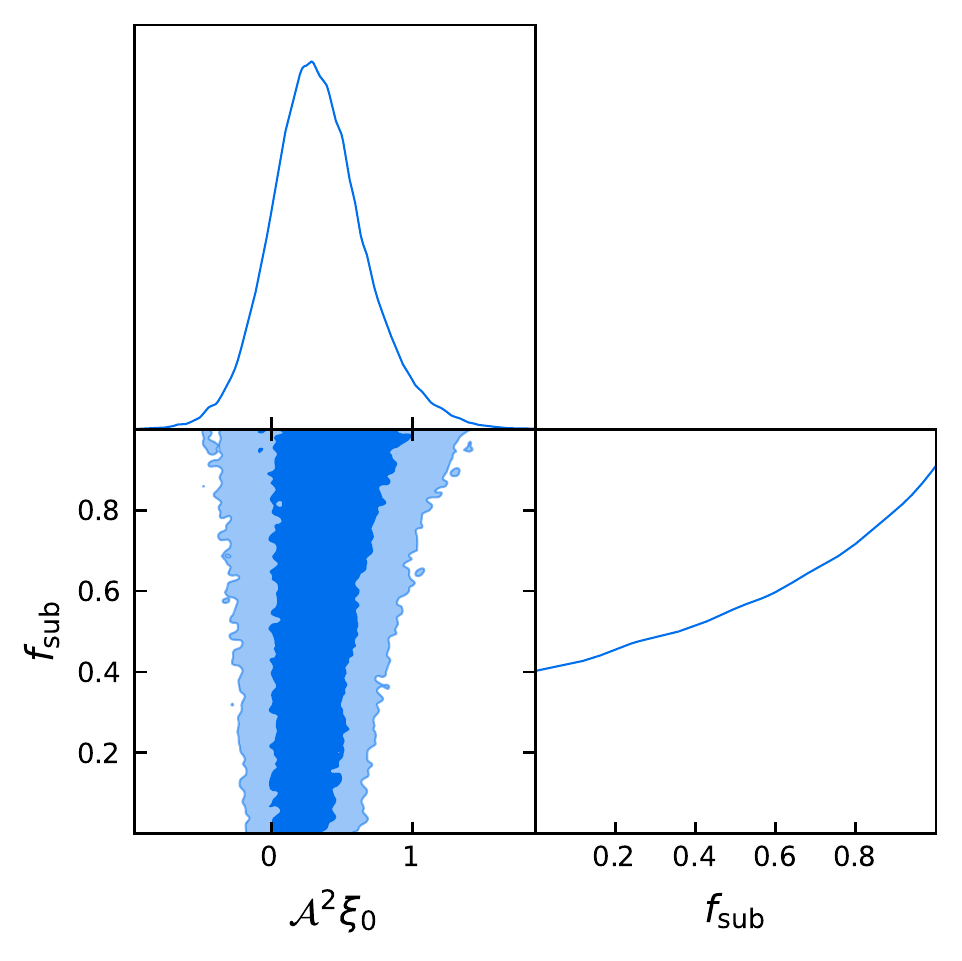}
    \caption{2D joint distribution and 1D marginalized distributions for the power spectrum normalization $\mathcal A^2\,\xi_0$ and the fraction $f_{\mathrm{sub}}$ of sub-Hubble scale string loops for Model II using data from Ref.~\cite{planck_constraint_2017}.}
    \label{fig:triangle_model2}
\end{figure}

We present in \reffig{triangle_model2} the results of MCMC sampling for the parameters in the more realistic Model II. Setting $\zeta_{\mathrm{min}}=0.1$ and $\zeta_{\mathrm{max}}=1$ as in \reffig{snr_model2}, we report an upper bound of $\mathcal A^2\,\xi_0<0.93$ at the 95\% confidence level.

Both \reffig{triangle_model1} and \reffig{triangle_model2} show that the data is consistent with the absence of a network of cosmic string loops. The model parameter $\zeta_0$ or $f_{\mathrm{sub}}$ is poorly constrained by the data due to the absence of a signal.

\section{Conclusion}\label{sec:conclusion}

We have summarized three different constructions of the quadratic estimator for direction dependent rotation of CMB polarization---the HO estimator \cite{Hu_2002}, the GMV estimator \cite{Maniyar_2021}, and one that is derived from the maximum likelihood estimator.
We have shown that the latter two are equivalent in the context of cosmic birefringence, and the first two are equivalent if temperature anisotropies are not used (a general sufficient condition for such equivalence is also stated). While the GMV estimator improves upon the HO estimator when both temperature and polarization anisotropies are taken into account, the improvement is numerically computed to be below 0.5\% on all relevant angular scales.

Although we have not seen a substantial reduction in the reconstruction noise by replacing the HO estimator with the GMV estimator, it might still be preferable to adopt the former for two reasons. First, the GMV estimator (as well as its reconstruction noise spectrum) is computed from a single integral (see \refeq{GMV} and \refeq{noise_GMV}), as opposed to one integral for every pair of anisotropy observables. Second, as pointed out in Ref.~\cite{Maniyar_2021}, the weights $\vect\Xi$ used in the GMV estimator are separable into sums of products of functions of $\vect l_1$ and functions of $\vect l_2$, allowing efficient computation via Fast Fourier Transform (FFT) without any further approximations (such as by setting $\tilde C_l^{TE} = 0$ in some formulae).

We present the reconstruction noise spectra of the HO estimator for the Planck mission, the Simons Observatory, as well as CMB-S4. This is compared to the reconstruction noise spectra of the single-pair $EB$ and $TB$ estimators \cite{Yadav_2009}. See \reffig{noise_curves}.

We have made SNR forecasts for phenomenological string network models (see \reffig{snr_model1} and \reffig{snr_model2}). It is found that contributions to the SNR are dominated by the reconstruction of modes on large angular scales: $L\lesssim 30$ for Planck SMICA, and $L\lesssim 100$ for the Simons Observatory and CMB-S4. When a continuous distribution of string loop sizes is accounted for, detectability does not strongly depend on the fraction of sub-horizon string loop sizes. In \reffig{snr_param_space}, it can be seen that existing Planck data should provide tight constraints in a significant portion of the theoretically plausible parameter space at $\mathcal A^2\,\xi_0 \gtrsim 1$.

The birefringence power spectrum extracted from Planck 2015 data \cite{planck_constraint_2017} has allowed us to place a constraint on $\mathcal A^2\,\xi_0$ using either of the string network models. Model I, which postulates circular string loops of identical radii, leads to an upper bound of $\mathcal A^2\,\xi_0 < 8.0$ at the 95\% confidence level, while the more realistic Model II, which postulates circular string loops of a continuous distribution of radii, leads to a stronger upper bound of $\mathcal A^2\,\xi_0 < 0.93$ at the 95\% confidence level. Model I leads to a weaker constraint because much of the parameter space describes a Universe populated by sub-Hubble scale string loops ($\zeta_0<1$), and the signal in the power spectrum is suppressed by an overall factor $\zeta_0$ \cite{Jain_2021}. On the other hand, most of the parameter space in Model II describes a Universe with many Hubble-scale string loops, leading to the prediction of a stronger signal. Simulations have shown that less than 20\% of the axion string loops are below the Hubble scale, while the remaining strings wrap the simulation box \cite{gorghetto_2018}, interpreted as string loops at the Hubble scale in Model II. It is for this reason that we view Model II, along with its bound on $\mathcal A^2\,\xi_0$, as a more accurate description of the phenomenology of cosmic string networks. In the future, more detailed numerical simulations of cosmic string network formation should provide more accurate models of the rotation field. The analysis presented in this work can also be performed for other models of cosmic birefringence, such as ALP domain wall scenarios presented in Ref.~\cite{Takahashi_Yin_2021}.

Third- and fourth-generation CMB experiments from the ground are expected to improve upon the SNR of the Planck mission by another one and two orders of magnitude, respectively, and will be able to either discover or rule out such cosmic string networks for most of the theoretically favored region of the parameter space (see \reffig{snr_param_space}).

We have justified the simplifying assumption that for estimating polarization rotation the lensed temperature and polarization anisotropies may be taken to be Gaussian random fields, by showing that the lensing potential $\phi$, which introduces mode couplings to the anisotropies between different Fourier modes, cause no bias at $\mathcal O(\phi)$ in estimating rotation.

While the quadratic estimators studied in this work can statistically detect a network of strings, it is sensitive to the combination $\mathcal A^2\,\xi_0$. The shape of the rotation power spectrum is expected to be nearly universal for $L \lesssim 50$, whether there are many small string loops or fewer larger loops \cite{Jain_2021}. Breaking the degeneracy between $\mathcal A$ and $\xi_0$ is crucial if one wishes to measure the axion-photon anomaly coefficient $\mathcal A$, whose value will have profound implications for particle physics at high energies as explained in \refsec{introduction}. This is in principle possible by measuring the discontinuity in the rotation angle across individual strings \cite{Agrawal_2019CMBmillikan}. Non-Gaussian statistics of the rotation field are expected to hold this information, which quadratic estimators are ill-equipped to probe. Moreover, linear stringy features imply phase correlations between Fourier modes of the rotation field, the confirmation of which will provide decisive evidence for topological defects as the origin of the ALP field. Extracting this information calls for further development of non-Gaussian statistical tests for polarization rotations in the CMB, or the employment of neural-network-based reconstruction techniques \cite{guzman2021reconstructing}.

\section*{Acknowledgments}

We thank Prateek Agrawal, Anson Hook and Junwu Huang for suggesting this research, and for their valuable input during both the early stage and the completion of this work. LD acknowledges the research grant support from the Alfred P.
Sloan Foundation (award number FG-2021-16495). SF~is supported by the Physics Division of Lawrence Berkeley National Laboratory.


\bibliographystyle{JHEP}
\bibliography{references}

\providecommand{\href}[2]{#2}\begingroup\raggedright\begin{thebibliography}{10}

\bibitem{Carroll_1998}
S.~M. Carroll, \emph{Quintessence and the rest of the world: Suppressing
  long-range interactions},
  \href{https://doi.org/10.1103/PhysRevLett.81.3067}{\emph{Phys. Rev. Lett.}
  {\bfseries 81} (1998) 3067}.

\bibitem{CarrollFieldJackiw_1990}
S.~M. Carroll, G.~B. Field and R.~Jackiw, \emph{Limits on a lorentz- and
  parity-violating modification of electrodynamics},
  \href{https://doi.org/10.1103/PhysRevD.41.1231}{\emph{Phys. Rev. D}
  {\bfseries 41} (1990) 1231}.

\bibitem{HarariSikivie_1992}
D.~{Harari} and P.~{Sikivie}, \emph{{Effects of a Nambu-Goldstone boson on the
  polarization of radio galaxies and the cosmic microwave background}},
  \href{https://doi.org/10.1016/0370-2693(92)91363-E}{\emph{Physics Letters B}
  {\bfseries 289} (1992) 67}.

\bibitem{SvrcekWitten_2006_StringTheory}
P.~{Svrcek} and E.~{Witten}, \emph{{Axions in string theory}},
  \href{https://doi.org/10.1088/1126-6708/2006/06/051}{\emph{Journal of High
  Energy Physics} {\bfseries 2006} (2006) 051}
  [\href{https://arxiv.org/abs/hep-th/0605206}{{\ttfamily hep-th/0605206}}].

\bibitem{Hui_2017}
L.~Hui, J.~P. Ostriker, S.~Tremaine and E.~Witten, \emph{Ultralight scalars as
  cosmological dark matter},
  \href{https://doi.org/10.1103/physrevd.95.043541}{\emph{Physical Review D}
  {\bfseries 95} (2017) }.

\bibitem{Kamionkowski2014axiverse}
M.~{Kamionkowski}, J.~{Pradler} and D.~G.~E. {Walker}, \emph{{Dark Energy from
  the String Axiverse}},
  \href{https://doi.org/10.1103/PhysRevLett.113.251302}{\emph{\prl} {\bfseries
  113} (2014) 251302} [\href{https://arxiv.org/abs/1409.0549}{{\ttfamily
  1409.0549}}].

\bibitem{Razieh2016axiverse}
R.~{Emami}, D.~{Grin}, J.~{Pradler}, A.~{Raccanelli} and M.~{Kamionkowski},
  \emph{{Cosmological tests of an axiverse-inspired quintessence field}},
  \href{https://doi.org/10.1103/PhysRevD.93.123005}{\emph{\prd} {\bfseries 93}
  (2016) 123005} [\href{https://arxiv.org/abs/1603.04851}{{\ttfamily
  1603.04851}}].

\bibitem{Agrawal_2019CMBmillikan}
P.~Agrawal, A.~Hook and J.~Huang, \emph{{A CMB Millikan experiment with cosmic
  axiverse strings}},
  \href{https://doi.org/10.1007/JHEP07(2020)138}{\emph{JHEP} {\bfseries 07}
  (2020) 138} [\href{https://arxiv.org/abs/1912.02823}{{\ttfamily
  1912.02823}}].

\bibitem{Hooft1980}
G.~Hooft, \emph{Naturalness, chiral symmetry, and spontaneous chiral symmetry
  breaking},  in \emph{Recent Developments in Gauge Theories}, G.~Hooft,
  C.~Itzykson, A.~Jaffe, H.~Lehmann, P.~K. Mitter, I.~M. Singer et~al., eds.,
  (Boston, MA), pp.~135--157, Springer US, (1980),
  \href{https://doi.org/10.1007/978-1-4684-7571-5_9}{DOI}.

\bibitem{Witten:1979ey}
E.~Witten, \emph{{Dyons of Charge e theta/2 pi}},
  \href{https://doi.org/10.1016/0370-2693(79)90838-4}{\emph{Phys. Lett. B}
  {\bfseries 86} (1979) 283}.

\bibitem{Minami_2020}
Y.~Minami and E.~Komatsu, \emph{New extraction of the cosmic birefringence from
  the planck 2018 polarization data},
  \href{https://doi.org/10.1103/physrevlett.125.221301}{\emph{Physical Review
  Letters} {\bfseries 125} (2020) }.

\bibitem{Fujita_2021}
T.~Fujita, K.~Murai, H.~Nakatsuka and S.~Tsujikawa, \emph{Detection of
  isotropic cosmic birefringence and its implications for axionlike particles
  including dark energy},
  \href{https://doi.org/10.1103/physrevd.103.043509}{\emph{Physical Review D}
  {\bfseries 103} (2021) }.

\bibitem{Kamionkowski_2009}
M.~Kamionkowski, \emph{How to derotate the cosmic microwave background
  polarization},
  \href{https://doi.org/10.1103/physrevlett.102.111302}{\emph{Physical Review
  Letters} {\bfseries 102} (2009) }.

\bibitem{Gluscevic2009Derotate}
V.~{Gluscevic}, M.~{Kamionkowski} and A.~{Cooray}, \emph{{Derotation of the
  cosmic microwave background polarization: Full-sky formalism}},
  \href{https://doi.org/10.1103/PhysRevD.80.023510}{\emph{\prd} {\bfseries 80}
  (2009) 023510} [\href{https://arxiv.org/abs/0905.1687}{{\ttfamily
  0905.1687}}].

\bibitem{Yadav_2009}
A.~P.~S. Yadav, R.~Biswas, M.~Su and M.~Zaldarriaga, \emph{Constraining a
  spatially dependent rotation of the cosmic microwave background
  polarization},
  \href{https://doi.org/10.1103/physrevd.79.123009}{\emph{Physical Review D}
  {\bfseries 79} (2009) }.

\bibitem{Gluscevic2012WMAP7}
V.~{Gluscevic}, D.~{Hanson}, M.~{Kamionkowski} and C.~M. {Hirata}, \emph{{First
  CMB constraints on direction-dependent cosmological birefringence from
  WMAP-7}}, \href{https://doi.org/10.1103/PhysRevD.86.103529}{\emph{\prd}
  {\bfseries 86} (2012) 103529}
  [\href{https://arxiv.org/abs/1206.5546}{{\ttfamily 1206.5546}}].

\bibitem{Jain_2021}
M.~Jain, A.~J. Long and M.~A. Amin, \emph{Cmb birefringence from
  ultralight-axion string networks},
  \href{https://doi.org/10.1088/1475-7516/2021/05/055}{\emph{Journal of
  Cosmology and Astroparticle Physics} {\bfseries 2021} (2021) 055}.

\bibitem{Hu_2002}
W.~Hu and T.~Okamoto, \emph{Mass reconstruction with cosmic microwave
  background polarization}, \href{https://doi.org/10.1086/341110}{\emph{The
  Astrophysical Journal} {\bfseries 574} (2002) 566–574}.

\bibitem{Maniyar_2021}
A.~S. Maniyar, Y.~Ali-Haïmoud, J.~Carron, A.~Lewis and M.~S. Madhavacheril,
  \emph{Quadratic estimators for cmb weak lensing},
  \href{https://doi.org/10.1103/physrevd.103.083524}{\emph{Physical Review D}
  {\bfseries 103} (2021) }.

\bibitem{planck_constraint_2017}
D.~Contreras, P.~Boubel and D.~Scott, \emph{Constraints on direction-dependent
  cosmic birefringence fromplanckpolarization data},
  \href{https://doi.org/10.1088/1475-7516/2017/12/046}{\emph{Journal of
  Cosmology and Astroparticle Physics} {\bfseries 2017} (2017) 046–046}.

\bibitem{Contreras_2017}
D.~Contreras, P.~Boubel and D.~Scott, \emph{Constraints on direction-dependent
  cosmic birefringence fromplanckpolarization data},
  \href{https://doi.org/10.1088/1475-7516/2017/12/046}{\emph{Journal of
  Cosmology and Astroparticle Physics} {\bfseries 2017} (2017) 046–046}.

\bibitem{Seljak:1998md}
U.~Seljak and M.~Zaldarriaga, \emph{{Polarization of microwave background:
  Statistical and physical properties}},  in \emph{{33rd Rencontres de Moriond:
  Fundamental Parameters in Cosmology}}, pp.~107--116, 1998,
  \href{https://arxiv.org/abs/astro-ph/9805010}{{\ttfamily astro-ph/9805010}}.

\bibitem{sherwin2021cosmic}
B.~D. Sherwin and T.~Namikawa, \emph{Cosmic birefringence tomography and
  calibration-independence with reionization signals in the cmb},  2021.

\bibitem{Hinshaw_2013}
G.~Hinshaw, D.~Larson, E.~Komatsu, D.~N. Spergel, C.~L. Bennett, J.~Dunkley
  et~al., \emph{Nine-year wilkinson microwave anisotropy probe ( wmap )
  observations: Cosmological parameter results},
  \href{https://doi.org/10.1088/0067-0049/208/2/19}{\emph{The Astrophysical
  Journal Supplement Series} {\bfseries 208} (2013) 19}.

\bibitem{Planck_2016}
N.~Aghanim, M.~Ashdown, J.~Aumont, C.~Baccigalupi, M.~Ballardini, A.~J. Banday
  et~al., \emph{Planckintermediate results},
  \href{https://doi.org/10.1051/0004-6361/201629018}{\emph{Astronomy \&
  Astrophysics} {\bfseries 596} (2016) A110}.

\bibitem{Polarbear_2020}
S.~Adachi, M.~A.~O. Aguilar~Faúndez, K.~Arnold, C.~Baccigalupi, D.~Barron,
  D.~Beck et~al., \emph{A measurement of the degree-scale cmb b-mode angular
  power spectrum with polarbear},
  \href{https://doi.org/10.3847/1538-4357/ab8f24}{\emph{The Astrophysical
  Journal} {\bfseries 897} (2020) 55}.

\bibitem{Bianchini_2020}
F.~Bianchini, W.~L.~K. Wu, P.~A.~R. Ade, A.~J. Anderson, J.~E. Austermann,
  J.~S. Avva et~al., \emph{Searching for anisotropic cosmic birefringence with
  polarization data from sptpol},
  \href{https://doi.org/10.1103/physrevd.102.083504}{\emph{Physical Review D}
  {\bfseries 102} (2020) }.

\bibitem{Namikawa_2020}
T.~Namikawa, Y.~Guan, O.~Darwish, B.~D. Sherwin, S.~Aiola, N.~Battaglia et~al.,
  \emph{Atacama cosmology telescope: Constraints on cosmic birefringence},
  \href{https://doi.org/10.1103/physrevd.101.083527}{\emph{Physical Review D}
  {\bfseries 101} (2020) }.

\bibitem{Choi_2020}
S.~K. Choi, M.~Hasselfield, S.-P.~P. Ho, B.~Koopman, M.~Lungu, M.~H. Abitbol
  et~al., \emph{The atacama cosmology telescope: a measurement of the cosmic
  microwave background power spectra at 98 and 150 ghz},
  \href{https://doi.org/10.1088/1475-7516/2020/12/045}{\emph{Journal of
  Cosmology and Astroparticle Physics} {\bfseries 2020} (2020) 045–045}.

\bibitem{Hirata_2003weaklensing}
C.~M. {Hirata} and U.~{Seljak}, \emph{{Analyzing weak lensing of the cosmic
  microwave background using the likelihood function}},
  \href{https://doi.org/10.1103/PhysRevD.67.043001}{\emph{\prd} {\bfseries 67}
  (2003) 043001} [\href{https://arxiv.org/abs/astro-ph/0209489}{{\ttfamily
  astro-ph/0209489}}].

\bibitem{Hirata_2003}
C.~M. Hirata and U.~Seljak, \emph{Reconstruction of lensing from the cosmic
  microwave background polarization},
  \href{https://doi.org/10.1103/physrevd.68.083002}{\emph{Physical Review D}
  {\bfseries 68} (2003) }.

\bibitem{PhysRevLett.78.2054}
U.~Seljak and M.~Zaldarriaga, \emph{Signature of gravity waves in the
  polarization of the microwave background},
  \href{https://doi.org/10.1103/PhysRevLett.78.2054}{\emph{Phys. Rev. Lett.}
  {\bfseries 78} (1997) 2054}.

\bibitem{Zaldarriaga_1997}
M.~Zaldarriaga and U.~Seljak, \emph{All-sky analysis of polarization in the
  microwave background},
  \href{https://doi.org/10.1103/physrevd.55.1830}{\emph{Physical Review D}
  {\bfseries 55} (1997) 1830–1840}.

\bibitem{PhysRevLett.78.2058}
M.~Kamionkowski, A.~Kosowsky and A.~Stebbins, \emph{A probe of primordial
  gravity waves and vorticity},
  \href{https://doi.org/10.1103/PhysRevLett.78.2058}{\emph{Phys. Rev. Lett.}
  {\bfseries 78} (1997) 2058}.

\bibitem{LEWIS_2006}
A.~Lewis and A.~Challinor, \emph{Weak gravitational lensing of the cmb},
  \href{https://doi.org/10.1016/j.physrep.2006.03.002}{\emph{Physics Reports}
  {\bfseries 429} (2006) 1–65}.

\bibitem{Namikawa_2017}
T.~Namikawa, \emph{Testing parity-violating physics from cosmic rotation power
  reconstruction},
  \href{https://doi.org/10.1103/physrevd.95.043523}{\emph{Physical Review D}
  {\bfseries 95} (2017) }.

\bibitem{Lewis_2011}
A.~Lewis, A.~Challinor and D.~Hanson, \emph{The shape of the cmb lensing
  bispectrum},
  \href{https://doi.org/10.1088/1475-7516/2011/03/018}{\emph{Journal of
  Cosmology and Astroparticle Physics} {\bfseries 2011} (2011) 018–018}.

\bibitem{Fabbian_2019}
G.~Fabbian, A.~Lewis and D.~Beck, \emph{Cmb lensing reconstruction biases in
  cross-correlation with large-scale structure probes},
  \href{https://doi.org/10.1088/1475-7516/2019/10/057}{\emph{Journal of
  Cosmology and Astroparticle Physics} {\bfseries 2019} (2019) 057–057}.

\bibitem{CAMB}
A.~{Lewis} and A.~{Challinor}, \emph{{CAMB: Code for Anisotropies in the
  Microwave Background}},  Feb., 2011.

\bibitem{Planck_2018_vi}
{Planck Collaboration}, {Aghanim, N.}, {Akrami, Y.}, {Ashdown, M.}, {Aumont,
  J.}, {Baccigalupi, C.} et~al., \emph{Planck 2018 results - vi. cosmological
  parameters}, \href{https://doi.org/10.1051/0004-6361/201833910}{\emph{A\&A}
  {\bfseries 641} (2020) A6}.

\bibitem{Planck_2018_iv}
{Planck Collaboration}, {Akrami, Y.}, {Ashdown, M.}, {Aumont, J.},
  {Baccigalupi, C.}, {Ballardini, M.} et~al., \emph{Planck 2018 results - iv.
  diffuse component separation},
  \href{https://doi.org/10.1051/0004-6361/201833881}{\emph{A\&A} {\bfseries
  641} (2020) A4}.

\bibitem{Ade_2019}
P.~Ade, J.~Aguirre, Z.~Ahmed, S.~Aiola, A.~Ali, D.~Alonso et~al., \emph{The
  simons observatory: science goals and forecasts},
  \href{https://doi.org/10.1088/1475-7516/2019/02/056}{\emph{Journal of
  Cosmology and Astroparticle Physics} {\bfseries 2019} (2019) 056–056}.

\bibitem{abazajian2016cmbs4}
K.~N. Abazajian, P.~Adshead, Z.~Ahmed, S.~W. Allen, D.~Alonso, K.~S. Arnold
  et~al., \emph{Cmb-s4 science book, first edition},  2016.

\bibitem{Pogosian_2019}
L.~Pogosian, M.~Shimon, M.~Mewes and B.~Keating, \emph{Future cmb constraints
  on cosmic birefringence and implications for fundamental physics},
  \href{https://doi.org/10.1103/physrevd.100.023507}{\emph{Physical Review D}
  {\bfseries 100} (2019) }.

\bibitem{Ferraro_2018}
S.~Ferraro and J.~C. Hill, \emph{Bias to cmb lensing reconstruction from
  temperature anisotropies due to large-scale galaxy motions},
  \href{https://doi.org/10.1103/physrevd.97.023512}{\emph{Physical Review D}
  {\bfseries 97} (2018) }.

\bibitem{gorghetto_2018}
M.~Gorghetto, E.~Hardy and G.~Villadoro, \emph{Axions from strings: the
  attractive solution},
  \href{https://doi.org/10.1007/jhep07(2018)151}{\emph{Journal of High Energy
  Physics} {\bfseries 2018} (2018) }.

\bibitem{Takahashi_Yin_2021}
F.~Takahashi and W.~Yin, \emph{Kilobyte cosmic birefringence from alp domain
  walls}, \href{https://doi.org/10.1088/1475-7516/2021/04/007}{\emph{Journal of
  Cosmology and Astroparticle Physics} {\bfseries 2021} (2021) 007}.

\bibitem{guzman2021reconstructing}
E.~Guzman and J.~Meyers, \emph{Reconstructing cosmic polarization rotation with
  resunet-cmb},  2021.

\end{thebibliography}\endgroup

\appendix


\end{document}